\documentclass[final,3p,times]{elsarticle}

\usepackage{amssymb}
\usepackage{amsthm}
\usepackage{amsmath}
\usepackage{booktabs}
\usepackage{layout}

\usepackage{geometry}
\usepackage{multirow}
\usepackage{tabularx}
\usepackage{algorithm}
\usepackage{algpseudocode}

\journal{Elsevier}

\begin{document}

\begin{frontmatter}

    \title{Transfer-learned Kolosov-Muskhelishvili Informed Neural Networks for Fracture Mechanics}

    \author[1]{Shuwei Zhou\corref{cor1}}
    \ead{shuwei.zhou@ibf.rwth-aachen.de}
    \author[1]{Christian Häffner}
    \author[2,3]{Shuancheng Wang}
    \author[1]{Sophie Stebner}
    \author[4]{Zhen Liao}
    \author[3]{Bing Yang}
    \author[1]{Zhichao Wei}
    \author[1]{Sebastian Münstermann}

    \cortext[cor1]{Corresponding author: Shuwei Zhou.}


    \affiliation[1]{organization={Institute of Metal Forming, RWTH Aachen University},
        city={Aachen},
        postcode={52072},
        country={Germany}}

    \affiliation[2]{organization={School of Mechanical Engineering, Sichuan University of Science and Engineering},
        city={Yibin},
        postcode={644005},
        country={PR China}}

    \affiliation[3]{organization={State Key Laboratory of Rail Transit Vehicle System, Southwest Jiaotong University},
        city={Chengdu},
        postcode={610031},
        country={PR China}}

    \affiliation[4]{organization={School of Mechanical Engineering, Sichuan University},
        city={Chengdu},
        postcode={610065},
        country={PR China}}

    \begin{abstract}
        Physics-informed neural networks have been widely applied to solid mechanics problems. However, balancing the governing partial differential equations and boundary conditions remains challenging, particularly in fracture mechanics, where accurate predictions strongly depend on refined sampling near crack tips. To overcome these limitations, a Kolosov-Muskhelishvili informed neural network with Williams enrichment is developed in this study. Benefiting from the holomorphic representation, the governing equations are satisfied by construction, and only boundary points are required for training. Across a series of benchmark problems, the Kolosov-Muskhelishvili informed neural network shows excellent agreement with analytical and finite element method references, achieving average relative errors below 1\% and $R^2$ above 0.99 for both mode~I and mode~II loadings. Furthermore, three crack propagation criteria (maximum tangential stress, maximum energy release rate, and principle of local symmetry) are integrated into the framework using a transfer learning strategy to predict crack propagation directions. The predicted paths are nearly identical across all criteria, and the transfer learning strategy reduces the required training time by more than 70\%. Overall, the developed framework provides a unified, mesh-free, and physically consistent approach for accurate and efficient crack propagation analysis.
    \end{abstract}




\end{frontmatter}

\section{Introduction}
\label{sec:intro}

Fracture mechanics in engineering components and materials has been a critical research area in solid mechanics for decades \cite{Wei.2024}. Owing to the lack of closed-form analytical solutions for complex crack problems, numerical methods have become the predominant approach. The finite element method (FEM) \cite{Wei.2022,Zhou.2026,Wei.2025}, phase-field methods (PFM) \cite{Ge.2024,Lo.2019,Fathi.2025}, and peridynamics (PD) \cite{Liu.2025,Xin.2025,Duan.2025}, among others \cite{Spada.2025, Yang.2021b}, are well-established tools for modeling crack initiation and propagation, and have achieved remarkable success across application domains \cite{Wang.2024g}. These approaches solve the governing partial differential equations (PDEs) for equilibrium and kinematics, coupled with constitutive models. Nonetheless, mesh-objective predictions for crack paths and energy release rates remain challenging: near-tip singularities often require specialized elements or strong local refinement in FEM; phase-field formulations introduce an internal length scale that interacts with the mesh; and PD solutions depend on the discretization horizon interplay \cite{Mao.2022}.

With growing computational resources, machine learning (ML) and deep learning (DL) have shown strong potential for PDE-governed solid mechanics \cite{Mirzaei.2025,Zhou.2024c}. They have been applied to surrogate modeling of constitutive behavior \cite{Pagan.2022,Sim.2025}, fatigue life prediction \cite{Peng.2022,Liu.2023e}, microstructure reconstruction \cite{Henrich.2020,Liang.2023}, discovery of closed-form surrogates via symbolic regression \cite{Zhou.2023b,Wang.2024j}, and identification of unknown material parameters \cite{Fehlemann.2025,Kong.2025}. However, ML/DL methods are intrinsically data-driven and typically demand sizable, high-quality datasets \cite{Lee.2024}. Although small-data strategies exist \cite{Keijzer.2004,Zhang.2024b,Xu.2023b}, sparse and noisy data can severely degrade performance and limit applicability \cite{Zhang.2024d}.

Physics-informed neural networks (PINNs) \cite{Raissi.2019} address data scarcity by embedding physics through loss terms that penalize the governing PDE residuals and boundary conditions. Since their introduction, PINNs have been applied to elasticity \cite{Xiong.2025,Luo.2024}, plasticity \cite{Weng.2025}, damage \cite{Hu.2024c}, and fatigue \cite{Jiang.2024,Feng.2024,Feng.2025,Zhou.2025}. In fracture mechanics, two main formulations have emerged: (i) energy-based approaches that minimize a variational functional, and (ii) PDE-based approaches that penalize field equations and boundary data. Representative energy-based studies include the variational PINN for phase-field brittle fracture by Goswami et al. \cite{Goswami.2020}, a physics-informed variational DeepONet for crack-path prediction \cite{Goswami.2022}, and an energy-minimization loss for quasi brittle crack growth under complex loading \cite{Zheng.2022}. While such approaches mitigate the need to explicitly resolve displacement jumps \cite{Zhao.2025}, accurate recovery of stress intensity factors (SIFs) and $T$-stress remains challenging \cite{Kolditz.2024}, and high-accuracy volume quadrature with strong tip refinement is typically required. On the PDE side, Gu et al. \cite{Gu.2023} enriched PINNs with Williams' functions to better capture near-tip singular fields, with extensions to fatigue growth \cite{Chen.2024} and bimaterial cracks \cite{Gu.2024}. Alternatively, PINNs based on holomorphic representation eliminate PDE-residual terms from the loss by satisfying the field equations a priori \cite{Calafa.2024}, and can be combined with enrichment to capture near-tip singularities \cite{Calafa.2025}. Nevertheless, performance can still be sensitive to the relative weighting of displacement, traction, and interface terms, especially when magnitudes differ across materials and load levels. Additionally, holomorphic networks can still not simulate crack growth automatically. In contrast, PDE-based PINNs remain sensitive to the balance between PDE and boundary condition (BC) losses and often require careful calibration.

To overcome these limitations, a Kolosov-Muskhelishvili informed neural network (KMINN) with Williams enrichment is developed for linear-elastic fracture mechanics (LEFM) in this study. Leveraging the KM holomorphic representation, the governing equations are satisfied by construction, and only boundary points are required for training. As a result, refined meshes near crack tips become unnecessary, while the enrichment accurately captures the singular near-tip fields. A root mean square (RMS) normalized boundary loss is used with physically consistent scaling to balance displacement, traction, and interface terms, rendering the training insensitive to load amplitude and material constants in practice. In addition, three crack-growth criteria, maximum tangential stress (MTS), maximum energy release rate (MERR), and the principle of local symmetry (PLS), are integrated within a transfer learning (TL) strategy to predict propagation directions and enhance training stability.

The remainder of this study is organized as follows. Section~\ref{sec:problem_formulation} shows the governing equations and the first-order Williams model expansion. Section~\ref{sec:methodology} gives the details of the KMINN with Williams enrichment framework. Section~\ref{sec:case_studies} provides case studies validating the approach. Section~\ref{sec:discussion} presents discussion and outlook, and Section~\ref{sec:conclusions} concludes the research.

\section{Problem Formulation}
\label{sec:problem_formulation}
This section presents the problem formulation of the KM potential for linear elasticity. Next, the first-order Williams model expansion is introduced. The two most important theories in this study are reviewed to make a more comprehensive understanding of the problem.
\subsection{Complex Representation of Linear Elasticity Equations}
\label{sec:KMINN_complex}
For isotropic, homogeneous, linear elastic materials, the strong form of the 2D equilibrium equation is given by,
\begin{equation}
    \frac{\partial \sigma_{11}}{\partial x_1} + \frac{\partial \sigma_{12}}{\partial x_2} + f_1 = 0,
    \label{eq:equilibrium_x}
\end{equation}
\begin{equation}
    \frac{\partial \sigma_{21}}{\partial x_1} + \frac{\partial \sigma_{22}}{\partial x_2} + f_2 = 0,
    \label{eq:equilibrium_y}
\end{equation}
where $\sigma_{ij}$ denotes the stress tensor, $f_i$ is the body force (zero in this study), and $x_i$ is the spatial coordinate. The stress tensor is related to the displacement field by Hooke's law,
\begin{equation}
    \sigma_{ij} = 2\mu \left( \varepsilon_{ij} + \frac{\nu}{1 - 2\nu} \varepsilon_{kk} \delta_{ij} \right), \quad i,j = 1,2,
    \label{eq:hooke}
\end{equation}
where $\varepsilon_{ij}$ is the strain tensor, $\nu$ represents the Poisson's ratio, and $\delta_{ij}$ is the Kronecker delta. $\mu$ is the shear modulus, and Lamé constant $\lambda$ is related to the Young's modulus $E$ and Poisson's ratio $\nu$ by the following equation,
\begin{equation}
    \mu = \frac{E}{2(1+\nu)}, \quad \lambda = \frac{E\nu}{(1+\nu)(1-2\nu)}.
    \label{eq:lam_mu}
\end{equation}

The strain tensor is related to the displacement field by the strain-displacement relation,
\begin{equation}
    \varepsilon_{ij} = \frac{1}{2} \left( \frac{\partial u_i}{\partial x_j} + \frac{\partial u_j}{\partial x_i} \right), \quad i,j = 1,2,
    \label{eq:strain_displacement}
\end{equation}
where $u_i$ is the displacement field. Moreover, the boundary conditions are given by,
\begin{equation}
    \left\{
    \begin{array}{ll}
        u_i = \bar{u}_i,             & \textrm{on} ~\Gamma_u \quad \textrm{(Dirichlet)}, \\[5pt]
        \sigma_{ij} n_j = \bar{t}_i, & \textrm{on} ~\Gamma_t \quad \textrm{(Neumann)},
    \end{array}
    \right.
\end{equation}
where $\bar{u}_i$ is the prescribed displacement, $\bar{t}_i$ is the prescribed traction, and $n_j$ is the outward normal vector on the boundary $\Gamma_u$ and $\Gamma_t$.

The complex representation of the linear elasticity equations is given by the complex potential function $\Phi(z)$, where $z = x + i y$ is the complex variable, and $\varphi(z)$ and $\psi(z)$ are analytic (holomorphic) functions representing the complex potentials on the domain. The Eqs.~\eqref{eq:equilibrium_x}, \eqref{eq:equilibrium_y}, \eqref{eq:hooke} and \eqref{eq:strain_displacement} can be described as Kolosov-Muskhelishvili (KM) formulations as follows,

\begin{equation}
    \left\{
    \begin{aligned}
         & \sigma_{xx} + \sigma_{yy} = 4\operatorname{Re} \left( \bar{z} \, \varphi''(z) + \psi'(z) \right), \\
         & \sigma_{xx} - \sigma_{yy} + 2i\sigma_{xy} = 2 \left( \bar{z} \, \varphi''(z) + \psi'(z) \right).
    \end{aligned}
    \right.
    \label{eq:KM_formulations_complex_original}
\end{equation}

Further, each stress and displacement component is given by,

\begin{equation}
    \left\{
    \begin{array}{l}
        \displaystyle \sigma_{xx} = \operatorname{Re} \left( 2\varphi' - \bar{z} \varphi'' - \psi' \right),                  \\[5pt]
        \displaystyle \sigma_{yy} = \operatorname{Re} \left( 2\varphi' + \bar{z} \varphi'' + \psi' \right),                  \\[5pt]
        \displaystyle \sigma_{xy} = \operatorname{Im} \left( \bar{z} \varphi'' + \psi' \right),                              \\[5pt]
        \displaystyle u_x = \dfrac{1}{2\mu} \operatorname{Re} \left( \kappa \varphi - \bar{z} \varphi' - \bar{\psi} \right), \\[5pt]
        \displaystyle u_y = \dfrac{1}{2\mu} \operatorname{Im} \left( \kappa \varphi - \bar{z} \varphi' - \bar{\psi} \right),
    \end{array}
    \right.
    \label{eq:KM_formulations}
\end{equation}
where $\kappa$ is Kolosov constant, defined as $\kappa = {(3 - \nu)} / {(1 + \nu)}$ for plane stress problem and $\kappa = 3 - 4\nu$ for plane strain problem, $\overline{(\cdot)}$ denotes the complex conjugate, and $\varphi'$ and $\varphi''$ represent the first and second order derivatives of the complex potential function $\varphi(z)$ concerning $z$, respectively.

\subsection{Williams Model Expansion} \label{sec:Williams_model}
The crack tip stress field of the Williams model is expressed in terms of the complex potentials $\varphi(z)$ and $\psi(z)$, which are holomorphic functions defined in the complex plane. The stress components can be expressed as,
\begin{equation}
    \begin{aligned}
        \sigma_{xx} - \sigma_{yy} + 2i\sigma_{xy} = 2 \left( \bar{z} \, \varphi''(z) + \psi'(z) \right)
        = A\,z^{-1/2}+B\,z^{-3/2}\bar z + C z^{0}.
        \label{eq:williams_stress_field}
    \end{aligned}
\end{equation}
where $A=-B={K_I}/{2\sqrt{2\pi}}, C=-T$ for mode $\mathrm{I}$, and $A=B={iK_{II}}/{2\sqrt{2\pi}}, C=-T$ for mode $\mathrm{II}$. The stress intensity factors $K_I$ and $K_{II}$ are the mode $\mathrm{I}$ and mode $\mathrm{II}$ stress intensity factors, respectively. The term $T$ is the T-stress, which could be ignored in the first-order Williams model. Thus, the crack tip stress field for mode $\mathrm{I}$ is expressed in terms of the stress intensity factor $K_I$ and the polar coordinates $(r, \theta)$ as follows,
\begin{equation}
    \left\{
    \begin{aligned}
        \sigma_{xx} & = \frac{K_I}{\sqrt{2\pi r}} \cos\frac{\theta}{2} \left( 1 - \sin\frac{\theta}{2} \sin\frac{3\theta}{2} \right), \\[5pt]
        \sigma_{yy} & = \frac{K_I}{\sqrt{2\pi r}} \cos\frac{\theta}{2} \left( 1 + \sin\frac{\theta}{2} \sin\frac{3\theta}{2} \right), \\[5pt]
        \sigma_{xy} & = \frac{K_I}{\sqrt{2\pi r}} \cos\frac{\theta}{2} \sin\frac{\theta}{2} \cos\frac{3\theta}{2},
    \end{aligned}
    \right.
    \label{eq:williams_stress}
\end{equation}
where $r$ is the distance from the crack tip, and $\theta$ is the angle measured from the crack plane. The angle $\theta$ is defined in the domain $\theta \in (-\pi, \pi]$, where $\theta = \pi$ and $\theta = -\pi$ denote the top and bottom crack surfaces, respectively. Additionally, the displacement field near the crack tip can be expressed as,
\begin{equation}
    \left\{
    \begin{aligned}
        u_x & = \frac{K_I}{2\mu} \sqrt{\frac{r}{2\pi}} \cos\frac{\theta}{2}
        \left[ \kappa - 1 + 2\sin^2\left( \frac{\theta}{2} \right) \right], \\[5pt]
        u_y & = \frac{K_I}{2\mu} \sqrt{\frac{r}{2\pi}} \sin\frac{\theta}{2}
        \left[ \kappa + 1 - 2\cos^2\left( \frac{\theta}{2} \right) \right].
    \end{aligned}
    \right.
    \label{eq:williams_displacement}
\end{equation}
Moreover, for mode $\mathrm{II}$, the stress field is given by,
\begin{equation}
    \left\{
    \begin{aligned}
        \sigma_{xx} & = \frac{K_{II}}{\sqrt{2\pi r}} \sin\frac{\theta}{2} \left( 1 + \cos\frac{\theta}{2} \sin\frac{3\theta}{2} \right),  \\[5pt]
        \sigma_{yy} & = -\frac{K_{II}}{\sqrt{2\pi r}} \sin\frac{\theta}{2} \left( 1 - \cos\frac{\theta}{2} \sin\frac{3\theta}{2} \right), \\[5pt]
        \sigma_{xy} & = -\frac{K_{II}}{\sqrt{2\pi r}} \sin^2\frac{\theta}{2} \cos\frac{3\theta}{2}.
    \end{aligned}
    \right.
    \label{eq:williams_stress_ii}
\end{equation}

The displacement field near the crack tip for mode $\mathrm{II}$ can be expressed as,
\begin{equation}
    \left\{
    \begin{aligned}
        u_x & = \frac{K_{II}}{2\mu} \sqrt{\frac{r}{2\pi}} \sin\frac{\theta}{2}
        \left[ \kappa - 1 + 2\cos^2\left( \frac{\theta}{2} \right) \right],     \\[5pt]
        u_y & = -\frac{K_{II}}{2\mu} \sqrt{\frac{r}{2\pi}} \cos\frac{\theta}{2}
        \left[ \kappa + 1 - 2\sin^2\left( \frac{\theta}{2} \right) \right].
    \end{aligned}
    \right.
\end{equation}

\section{Methodology}
\label{sec:methodology}
The methodology of the KMINN with Williams enrichment is presented in this section.
First, the KMINN with the Williams enrichment framework is built to capture the stress and displacement fields near the crack tip. An RMS normalized boundary loss is used with physically consistent scaling to balance displacement, and traction terms, rendering the training insensitive to load amplitude and material constants in practice. Next, a precise analytic method is introduced to estimate the SIFs from the KMINN output. Finally, three crack propagation criteria are integrated within a TL strategy to predict propagation directions for each step and enhance training stability.

\subsection{Kolosov-Muskhelishvili Informed Neural Networks (KMINN) with Williams Enrichment}\label{sec:KMINN_pih}
KMINN is a subclass of complex-valued neural networks that extends real-valued PINNs to the complex domain. The core component is a KM network that uses holomorphic activation functions. This design makes the network output holomorphic on the computational domain. This property is well-suited for 2D linear elasticity problems. As outlined in Section~\ref{sec:KMINN_complex}, a 2D linear elastic solution is fully described by the two holomorphic KM potentials. By directly learning these potentials with a KM network, KMINN satisfies the governing equations by construction \cite{Calafa.2025}. Based on this, the KMINN with Williams enrichment is developed in this study to capture the stress field singularities near the crack tip by requiring only BC collocation points.

\begin{figure*}[h]
    \centering
    \includegraphics[width=0.88\textwidth]{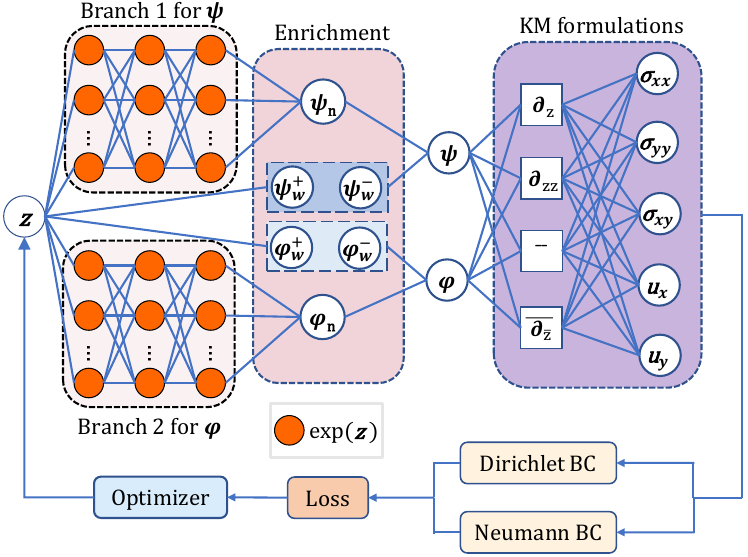}
    \caption{KMINN framework with Williams enrichment for each subdomain.}
    \label{fig:KMINN_framework_williams}
\end{figure*}

For each subdomain, two branches are used to approximate the holomorphic KM potentials, denoted by $\varphi_n$ and $\psi_n$ as shown in Fig.~\ref{fig:KMINN_framework_williams}. A neural network that approximates a holomorphic function must be built from holomorphic components. Since a network is a composition of linear operations and activation functions, the activation functions must be complex differentiable. An entire function is employed as the activation function of the KMINNs as given by,
\begin{equation}
    f(z) = \exp(z).
    \label{eq:transcendental_entire_function}
\end{equation}
The choice of the entire activation is discussed in detail in \cite{Calafa.2024}. Note that the exponential equation has analytic and calculus readiness, caused by its derivatives being itself. Moreover, sums of exponentials form a rich basis for approximating holomorphic functions, which underpins the KMINN universal approximation property described for the framework. Whereas exponential activations can amplify magnitudes and gradients, an exponential-aware scaling \cite{Calafa.2024} and He's initialization \cite{He.2015} schemes are applied to stabilize forward and backward passes.

Specifically, for each complex linear layer $\mathcal{L}_\ell:\mathbb{C}^{n_{\mathrm{in}}^{(\ell)}}\!\to\!\mathbb{C}^{n_{\mathrm{out}}^{(\ell)}}$, the weights are initialized by a complex He's scheme with layerwise variance parameter $\rho_\ell$,
\begin{equation}
    \displaystyle \Re W^{(\ell)}_{ij},\,\Im W^{(\ell)}_{ij} \sim \mathcal{N}\!\Big(0,\ \tfrac{\rho_\ell}{2\,n_{\mathrm{in}}^{(\ell)}}\Big),
    \qquad b^{(\ell)}=0.
    \label{eq:he_complex}
\end{equation}
Let $x^{(0)}$ denote a representative batch of boundary inputs. A single forward sweep at initialization produces,
\begin{equation}
    y^{(\ell)} = W^{(\ell)} x^{(\ell-1)}, \qquad
    x^{(\ell)} = \phi\!\big(y^{(\ell)}\big)=\exp\!\big(y^{(\ell)}\big),
\end{equation}
and the empirical second moment $m^{(\ell)}=\widehat{\mathbb{E}}\!\left[|x^{(\ell)}|^2\right]$.
Given a parameter $\beta>0$ and the number of Gaussian pre-stabilizing layers $M_e$, the layerwise variance parameter $\rho_\ell$ is chosen as,
\begin{equation}
    \rho_\ell =
    \begin{cases}
        \displaystyle \frac{\beta}{\,\widehat{\mathbb{E}}\!\left[|x^{(\ell-1)}|^2\right]}\,, & \ell \le M_e, \\[10pt]
        \displaystyle \beta\,e^{-\beta},                                                     & \ell > M_e.
    \end{cases}
    \label{eq:rho_rule}
\end{equation}
Practically, we initialize layer $\ell$, propagate once to update the statistics for layer $\ell{+}1$, and continue. This data-dependent layerwise rule keeps the empirical second moment approximately constant through $\exp(\cdot)$, damping activation growth and stabilizing backpropagation from the first steps. In addition, small variance initialization and gradient clipping are applied to further prevent gradient explosion.

\begin{figure}[h]
    \centering
    \includegraphics[width=0.8\textwidth]{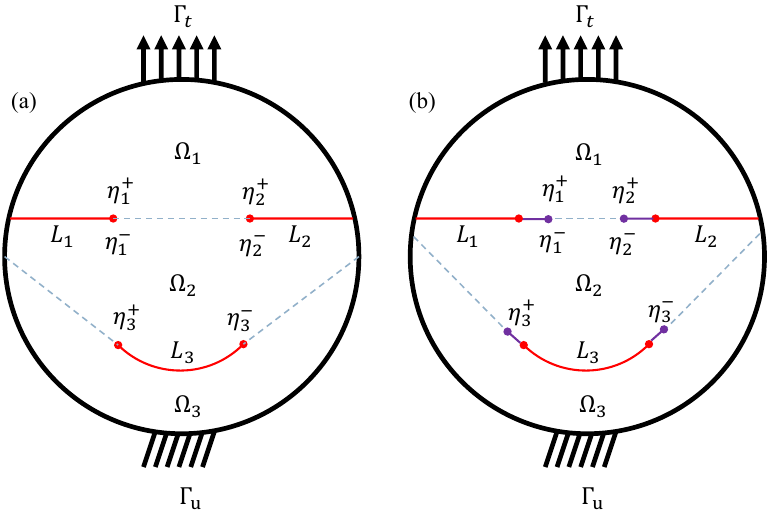}
    \caption{Domain decomposition for the KMINN with Williams enrichment. (a) Initial pre-crack domain. (b) Post-crack domain. Red solid line represents the initial crack line, purple solid line represents the crack line after crack propagation, and blue dashed line represents the interface between two subdomains to ensure the holomorphicity of Williams enrichment part.}
    \label{fig:domain_decomposition}
\end{figure}

Furthermore, the first-order Williams enrichment is applied to the KMINN to generate the Williams model potentials $\varphi^{\pm}_{W}$, $\psi^{\pm}_{W}$.
Based on the first-order Williams model described in Section~\ref{sec:Williams_model}, the stress field near the crack tip can be expressed in terms of the complex potentials $\varphi$ and $\psi$ as follows \cite{Calafa.2025},
\begin{equation}
    \left\{
    \begin{aligned}
        \varphi & = \frac{K}{\sqrt{2\pi}} \sqrt{z},                                      \\[5pt]
        \psi    & = \left( \bar{K} - \frac{K}{2} \right) \frac{1}{\sqrt{2\pi}} \sqrt{z},
    \end{aligned}
    \right.
    \label{eq:williams_complex_potentials}
\end{equation}
where $K = K_\mathrm{I} - i K_\mathrm{II}$ is the complex stress intensity factor, and $\bar{K} = K_I + i K_{II}$ is its complex conjugate. It should be mentioned that Eq.~\eqref{eq:williams_complex_potentials} is not entire over the full domain due to the $\sqrt{z}$ term. This term introduces a multi-valuedness that requires a branch cut to be defined. Thus, domain decomposition (DD) is employed to partition the domain along the crack line, ensuring the Williams enrichment potentials remain well-defined and single-valued within each subdomain. This partitioning also avoids impacting the training process during crack propagation. In this study, the DD is achieved through interface lines as shown in Fig.~\ref{fig:domain_decomposition}. In the initial configuration, the artificial interface is taken along the crack extension for straight cracks or along a tangent line for curved cracks; after each propagation step, the interface is updated to remain tangent to the latest crack propagation direction.

To fit the arbitrary position of the crack shape, the rotation and translation of the KM complex potentials are applied, which is expressed as,
\begin{equation}
    \left\{
    \begin{aligned}
        \varphi(z) & = e^{i\alpha} \, \hat{\varphi}\left(e^{-i\alpha}(z - \eta)\right),                                                                 \\[5pt]
        \psi(z)    & = e^{-i\alpha} \, \hat{\psi}\left(e^{-i\alpha}(z - \eta)\right) - \bar{\eta} \, \hat{\varphi}'\left(e^{-i\alpha}(z - \eta)\right),
    \end{aligned}
    \right.
\end{equation}
where $\hat{\varphi}$ and $\hat{\psi}$ are the KM complex potentials defined in the subdomain $\Omega_j$, $\alpha$ is the rotation angle, and $\eta$ is the translation vector.

In each subdomain $\Omega_j$, the complex potentials are,
\begin{equation}
    \left\{
    \begin{aligned}
        \varphi_j(z) & = \varphi_{n,j}(z) + \varphi^{+}_{W,j}(z) + \varphi^{-}_{W,j}(z), \\
        \psi_j(z)    & = \psi_{n,j}(z) + \psi^{+}_{W,j}(z) + \psi^{-}_{W,j}(z),
    \end{aligned}
    \right. \quad z \in \Omega_j,
    \label{eq:calafa_complex_potentials}
\end{equation}
where $\varphi_{n,j}(z)$ and $\psi_{n,j}(z)$ are the holomorphic potentials restricted to $\Omega_{j}$, and $\varphi^{+}_{W,j}(z)$, $\varphi^{-}_{W,j}(z)$, $\psi^{+}_{W,j}(z)$, and $\psi^{-}_{W,j}(z)$ are the Williams model potentials for the positive and negative crack tips in subdomain $\Omega_j$. The branch cut of $\sqrt{z}$ is aligned with the crack line to avoid phase jumps. Thus, combining Eqs.~\eqref{eq:williams_complex_potentials}--\eqref{eq:calafa_complex_potentials} gives,

\begin{equation}
    \left\{
    \begin{aligned}
        \varphi^{+}_{W,j}(z) & = \frac{K^{+}_{l,j}}{\sqrt{2\pi}}{e^{ia^{+}_{l,j}}} \sqrt{{e^{-ia^{+}_{l,j}}}(z-p^{+}_{l,j})}, \\
        \varphi^{-}_{W,j}(z) & = \frac{K^{-}_{l,j}}{\sqrt{2\pi}}{e^{ia^{-}_{l,j}}} \sqrt{{e^{-ia^{-}_{l,j}}}(z-p^{-}_{l,j})}, \\
        \psi^{+}_{W,j}(z)    & = \frac{1}{\sqrt{2\pi}}\Bigl[
        \bigl(\bar K^{+}_{l,j}-\tfrac{ K^{+}_{l,j}}{2}\bigr)\,
        e^{-i a^{+}_{l,j}} \sqrt{\,e^{-i a^{+}_{l,j}}\!\left(z-p^{+}_{l,j}\right)}
        - \frac{K^{+}_{l,j} \, \bar{p}^{+}_{l,j}}{2}\,
        \frac{1}{\sqrt{\,e^{-i a^{+}_{l,j}}\!\left(z-p^{+}_{l,j}\right)}}\Bigr],                                              \\
        \psi^{-}_{W,j}(z)    & = \frac{1}{\sqrt{2\pi}}\Bigl[
        \bigl(\bar K^{-}_{l,j}-\tfrac{ K^{-}_{l,j}}{2}\bigr)\,
        e^{-i a^{-}_{l,j}} \sqrt{\,e^{-i a^{-}_{l,j}}\!\left(z-p^{-}_{l,j}\right)}- \frac{K^{-}_{l,j} \, \bar{p}^{-}_{l,j}}{2}\,
        \frac{1}{\sqrt{\,e^{-i a^{-}_{l,j}}\!\left(z-p^{-}_{l,j}\right)}}\Bigr].
    \end{aligned}
    \right. \quad z \in \Omega_j,
    \label{eq:calafa_williams_complex_potentials}
\end{equation}

Thus, the KMINN output is the sum of the holomorphic part and the Williams enrichment part. Its holomorphicity in each subdomain is proven in \ref{sec:Holomorphicity_proof}.

The KM formulations in Eq.~\eqref{eq:KM_formulations} are applied to obtain the stress and displacement fields in each subdomain. Unlike classical PINNs, the KMINN loss function enforces only the BCs, as the governing PDEs are already satisfied automatically for holomorphic functions. The loss function is defined as a weighted sum of the mean squared errors on the boundaries,
\begin{equation}
    \mathcal{L}
    = \sum_{s\in\mathcal{S}} w_{\mathrm{len}}^{(s)}
    \Big( \mathcal{L}_u^{(s)} + \mathcal{L}_t^{(s)} + \mathcal{L}_{I}^{(s)} \Big),
    \qquad
    w_{\mathrm{len}}^{(s)}=\frac{l_s}{\sum_{s'} l_{s'}},
    \label{eq:KMINN_loss}
\end{equation}
where $\mathcal{L}_u^{(s)}$, $\mathcal{L}_t^{(s)}$, and $\mathcal{L}_{I}^{(s)}$ are the loss functions on the Dirichlet boundary $\Gamma_u$, Neumann boundary $\Gamma_t$, and interface boundary $\Gamma_I$ respectively. $w_{\mathrm{len}}^{(s)}$ is the weight of the boundary $s$ based on the length of the boundary. The interface condition between each subdomains is defined as \cite{Wang.2024m},
\begin{equation}
    \left\{
    \begin{aligned}
         & \llbracket\boldsymbol{\sigma}_I \cdot \boldsymbol{n} \rrbracket = 0, \\
         & \llbracket\boldsymbol{u}_I\rrbracket = 0,
    \end{aligned}
    \right.
    \label{eq:interface_condition}
\end{equation}
where $\boldsymbol{n}$ is the outward normal vector on the interface boundary, $\boldsymbol{\sigma}_I$ is the stress field on the interface boundary, and $\boldsymbol{u}_I$ is the displacement field on the interface boundary. Furthermore, pre-normalized fields are defined as follows,
\begin{equation}
    \tilde{\boldsymbol{\sigma}}=\frac{\boldsymbol{\sigma}}{\sigma_{\mathrm{ref}}},
    \qquad
    \tilde{\boldsymbol{u}}=\frac{\boldsymbol{u}}{u_{\mathrm{ref}}},\quad
    u_{\mathrm{ref}}=\frac{\sigma_{\mathrm{ref}} L_{\mathrm{ref}}}{E^{'}},
    \label{eq:KMINN_loss_normalization}
\end{equation}
where $\sigma_{\mathrm{ref}}$ is the RMS of applied tractions on Neumann boundaries, $L_{\mathrm{ref}}$ is a domain length scale. $E^{'}$ is the effective modulus, which is $E$ for plane stress and $E/(1-\nu^2)$ for plane strain. The method makes all boundary loss terms unit-free and comparable in magnitude across different BC types and across materials with different properties, improving the training stability and generalization ability.

The individual loss terms for the Dirichlet $\Gamma_u$, Neumann $\Gamma_t$, and interface $\Gamma_I$ boundaries are defined as,
\begin{equation}
    \left\{
    \begin{aligned}
         & \mathcal{L}_u^{(s)}
        = \frac{1}{N_u^{(s)}}\sum_{z_i\in\Gamma_u^{(s)}}
        \big\|\tilde{\boldsymbol{u}}(z_i)-\tilde{\bar{\boldsymbol{u}}}(z_i)\big\|_2^{2},              \\[4pt]
         & \mathcal{L}_t^{(s)}
        = \frac{1}{N_t^{(s)}}\sum_{z_i\in\Gamma_t^{(s)}}
        \big\|\tilde{\boldsymbol{\sigma}}(z_i)\boldsymbol{n}-\tilde{\boldsymbol{t}}(z_i)\big\|_2^{2}, \\[4pt]
         & \mathcal{L}_I^{(s)}
        = \frac{1}{N_I^{(s)}}\sum_{z_i\in\Gamma_I^{(s)}}
        \left(\big\|\llbracket\tilde{\boldsymbol{u}}\rrbracket(z_i)\big\|_2^{2}
        +\big\|\llbracket\tilde{\boldsymbol{\sigma}}\boldsymbol{n}\rrbracket(z_i)\big\|_2^{2}\right),
    \end{aligned}
    \right.
    \label{eq:loss_terms}
\end{equation}
where $N_u^{(s)}$, $N_t^{(s)}$, and $N_I^{(s)}$ are the number of training points on their respective boundaries. This approach is highly efficient as it only requires training points on the domain boundary, not its interior. Thus, KMINN does not require interior meshes nor PDE residual terms in the training process.

Referring to \cite{Wang.2024h}, the mixed training strategy is applied to train the KMINN with Williams enrichment to further improve the training stability and generalization ability. Simply, the KMINN is trained by Adam optimizer \cite{Kingma.2015} in the first stage, and then the L-BFGS optimizer \cite{Zhang.2026} is employed to fine-tune the model in the second stage. The strategy combines the advantages of fast convergence of Adam optimizer and the stability of L-BFGS optimizer.

\subsection{Stress Intensity Factor Estimation} \label{sec:sif_estimation}
In LEFM, the SIFs are fundamental parameters to characterize stress and displacement fields near the crack tip \cite{Abdolvand.2022}. The SIFs can be extracted from the network parameters like $K^{+}_1$, $K^{-}_1$, $K^{+}_2$, and $K^{-}_2$. However, this direct extraction is sensitive to the local Williams expansion and to how accurately the network resolves the $r^{-1/2}$ fields in the immediate vicinity of the crack tip, which can introduce fitting bias. To obtain more robust and globally consistent estimates of $K_I$ and $K_{II}$, an interaction-integral (also referred to as an $I$-integral) formulation is additionally utilized in this study.

Among a series of fracture methods, the J-integral method which was originally proposed by Rice \cite{Rice.1968} is the most prevalent for its clear physical meaning. J-integral method is based on the strain-energy release rate and its path-independence facilitates its numerical implementations. The J-integral is defined as,
\begin{equation}
    J \;=\; \oint_{\Gamma} \Big( W\,\delta_{1j} - \sigma_{ij} \, u_{i,1} \Big) n_j \,\mathrm d\Gamma,
    \label{eq:j_integral}
\end{equation}
where $W=\tfrac12\sigma_{ij}\varepsilon_{ij}$ is the strain energy density, $u_{i,1}\!=\!\partial u_i/\partial x_1$, and $n_j$ is the outward unit normal to the contour $\Gamma$.

Only isotropic elastic materials are considered in this study. The I-integral combines the actual field $(u, \sigma)$ with an analytically known auxiliary field $(u_i^{\text{aux}},\sigma_{ij}^{\text{aux}})$, which corresponds to a unit mode I or unit mode II singular solution. Therefore, the auxiliary stress and displacement fields are given by \cite{Yau.1980},
\begin{equation}
    \left\{
    \begin{aligned}
        u_i^{\text{aux}} = \sqrt{\frac{r}{2\pi}} \left[ K_I^{\text{aux}} f_i^{I}(\theta) + K_{II}^{\text{aux}} f_i^{II}(\theta) \right], \\
        \sigma_{ij}^{\text{aux}} = \frac{1}{\sqrt{2\pi r}} \left[ K_I^{\text{aux}} g_{ij}^{I}(\theta) + K_{II}^{\text{aux}} g_{ij}^{II}(\theta) \right],
    \end{aligned}
    \right.
    \label {eq:auxiliary_fields}
\end{equation}
where $K_I^{\text{aux}}$ and $K_{II}^{\text{aux}}$ are the auxiliary SIFs, and $f_i^{I}(\theta)$, $f_i^{II}(\theta)$, $g_{ij}^{I}(\theta)$, and $g_{ij}^{II}(\theta)$ are the auxiliary functions defined in the polar coordinates $(r, \theta)$, which can be expressed as,
\begin{equation}
    \left\{
    \begin{aligned}
        f_i^{I}(\theta)     & =
        \begin{pmatrix}
            \dfrac{\kappa - \cos\theta}{2\mu} \cos\frac{\theta}{2} \\[6pt]
            \dfrac{\kappa - \cos\theta}{2\mu} \sin\frac{\theta}{2}
        \end{pmatrix},                   \\[10pt]
        f_i^{II}(\theta)    & =
        \begin{pmatrix}
            \dfrac{2 + \kappa + \cos\theta}{2\mu} \sin\frac{\theta}{2} \\[6pt]
            \dfrac{2 - \kappa - \cos\theta}{2\mu} \cos\frac{\theta}{2}
        \end{pmatrix},               \\[10pt]
        g_{ij}^{I}(\theta)  & =
        \begin{pmatrix}
            \dfrac{3}{4} \cos\frac{\theta}{2} - \dfrac{1}{4} \cos\frac{5\theta}{2} &
            \dfrac{7}{4} \sin\frac{\theta}{2} - \dfrac{1}{4} \sin\frac{5\theta}{2}   \\[6pt]
            \dfrac{7}{4} \sin\frac{\theta}{2} - \dfrac{1}{4} \sin\frac{5\theta}{2} &
            \dfrac{5}{4} \cos\frac{\theta}{2} + \dfrac{1}{4} \cos\frac{5\theta}{2}
        \end{pmatrix}, \\[10pt]
        g_{ij}^{II}(\theta) & =
        \begin{pmatrix}
            \dfrac{7}{4} \sin\frac{\theta}{2} - \dfrac{1}{4} \sin\frac{5\theta}{2} &
            \dfrac{3}{4} \cos\frac{\theta}{2} - \dfrac{1}{4} \cos\frac{5\theta}{2}   \\[6pt]
            \dfrac{3}{4} \cos\frac{\theta}{2} - \dfrac{1}{4} \cos\frac{5\theta}{2} &
            -\dfrac{1}{4} \sin\frac{\theta}{2} + \dfrac{1}{4} \sin\frac{5\theta}{2}
        \end{pmatrix}.
    \end{aligned}
    \right.
    \label{eq:auxiliary_functions}
\end{equation}

For a two-dimensional crack in an isotropic linear-elastic body, the $J$-integral can be related to the mode~I and mode~II SIFs as \cite{Shih.1986},
\begin{equation}
    J = \frac{1}{E^{'}} \left( K_I^2 + K_{II}^2 \right),
    \label{eq:j_integral_sif}
\end{equation}
where $E^{'} = E$ in plane stress and $E^{'} = E/(1-\nu^2)$ in plane strain. Similarly, the I-integral can be calculated by,
\begin{equation}
    I = \frac{2}{E^{'}} \left( K_I K_I^{\text{aux}} + K_{II} K_{II}^{\text{aux}} \right),
    \label{eq:j_integral_sif_i}
\end{equation}
where $I$ is the mutual energy release rate, which is defined as the energy release rate per unit crack area. The set of SIF {$K^{aux}_I$, $K^{aux}_{II}$} can be determined by taking $\{1, 0\}$ or $\{0, 1\}$ for mode $\mathrm{I}$ or mode $\mathrm{II}$, respectively. The mode~I and mode~II SIFs then follow directly as,
\begin{equation}
    \left\{
    \begin{aligned}
        K_I    & = \frac{E^{'}}{2}{I^{(I)}},  \\[4pt]
        K_{II} & = \frac{E^{'}}{2}{I^{(II)}}.
    \end{aligned}
    \right.
    \label{eq:sif_estimation}
\end{equation}
In practice, the SIFs can be estimated using the I-integral method with circular contours from the auxiliary stress and displacement fields in the KMINN results. In theory, for linear elastic fracture without body forces, the interaction integral is path-independent. However, in practice, the SIFs results are influenced by boundary effects or near tip singularities due to crack length and tip position \cite{Paik.2021}.  Consequently, we compute $I$ over several contours of different radii and identify a radius range in which the resulting $K_I$ and $K_{II}$ are stable. This procedure is discussed below.

\subsection{Fracture Growth Criteria}
\label{sec:fracture_criteria}

For fracture modeling, if the crack propagates or not is determined by the fracture criteria. The SIF $K_I$ and $K_{II}$ are obtained from the KMINN by the I-integral method as discussed in Sec.~\ref{sec:sif_estimation}. Thus, the situation of crack propagation for mixed-mode loading is evaluated by equivalent strength criterion, which is given by,
\begin{equation}
    \left(\frac{K_I}{K_{I,~c}}\right)^2 + \left(\frac{K_{II}}{K_{II,~c}}\right)^2 \;\begin{cases}
        < 1, & \text{no growth}, \\
        = 1, & \text{onset},     \\
        > 1, & \text{growth},
    \end{cases}
    \label{eq:equivalent_strength_criterion}
\end{equation}
where $K_{I,~c}$ and $K_{II,~c}$ are the critical SIFs for mode $\mathrm{I}$ and mode $\mathrm{II}$, respectively.

To predict the crack growth direction and path, several fracture criteria are commonly used in engineering applications. In this section, the MTS, MERR, and the PLS criteria are introduced for achieving the crack growth direction prediction in KMINN.

The MTS criterion is a fracture criterion that is based on the maximum tangential stress at the crack tip, which was originally proposed by Erdogan and Sih \cite{Erdogan.1963}. The MTS criterion is widely used in engineering applications to predict the crack propagation direction of materials. The MTS criterion is implemented in the KMINN by calculating the maximum tangential stress at the crack tip and determining the crack growth direction based on the MTS. The near crack tip stress field under polar coordinates $(r, \theta)$ can be expressed as,

\begin{equation}
    \left\{
    \begin{aligned}
        \displaystyle \sigma_{\theta\theta}(r,\theta) & = \frac{1}{\sqrt{2\pi r}}
        \Big[\,K_I \cos^3\!\frac{\theta}{2}\;-\;3K_{II}\,\sin\frac{\theta}{2}\cos^2\!\frac{\theta}{2}\,\Big], \\[3pt]
        \displaystyle \tau_{r\theta}(r,\theta)        & = \frac{1}{\sqrt{2\pi r}}
        \Big[\,K_I \sin\frac{\theta}{2}\cos\frac{\theta}{2}\;+\;
            K_{II}\,\cos\frac{\theta}{2}\big(1-3\sin^2\frac{\theta}{2}\big)\,\Big].
    \end{aligned}
    \right.
    \label{eq:mts_stress_fixed}
\end{equation}

where $\sigma_{\theta\theta}$ is the normal stress, $\tau_{r\theta}$ is the shear stress, respectively. The principal direction should satisfy $\tau_{r\theta}(r\!\to\!0,\theta)=0$, among those directions the crack advances where $\sigma_{\theta\theta}$ reaches its maximum tensile value. The crack growth direction can be calculated by,
\begin{equation}
    \tau_{r\theta}=0,\qquad
    \frac{\partial \sigma_{\theta\theta}}{\partial \theta}=0,\qquad
    \sigma_{\theta\theta}>0.
    \label{eq:mts_condition}
\end{equation}

For a pure SIF-domain field, the crack growth angle $\theta_T$ can be calculated by,
\begin{equation}
    \theta_T= \arccos \left( \frac{3K_{II}^2 + \sqrt{K_I^4 + 8K_I^2 K_{II}^2}}{K_I^2 + 9K_{II}^2} \right).
    \label{eq:mts_crack_growth_angle}
\end{equation}

The MERR criterion is another fracture criterion that is based on the maximum energy release rate at the crack tip, which is an extension of Griffith's theory to mixed-mode loading \cite{Zhao.2024c}. It assumes that the crack propagates in the direction of the maximum energy release rate $G$. This is the path of least resistance from an energetic standpoint. The crack growth direction can be determined by the maximum energy release rate $G$ and the angle $\theta$,

\begin{equation}
    \frac{dG(\theta)}{d\theta} = 0, \quad \text{and} \quad \frac{d^2G(\theta)}{d\theta^2} < 0.
\end{equation}

For isotropic materials, the equation of $G(\theta)$ can be obtained by,
\begin{equation}
    G(\theta) = \frac{1}{E^{'}}(K_I^2 + K_{II}^2).
    \label{eq:merr}
\end{equation}

The crack growth angle $\theta_E$ is calculated by,
\begin{equation}
    \theta_E = \mathrm{argmax}_{\theta}\, \left(\frac{1}{E^{'}}(K_I^2 + K_{II}^2)\right).
    \label{eq:merr_crack_growth_angle}
\end{equation}

The PLS criterion is a widely accepted criterion for predicting the initiation angle of crack propagation under mixed-mode loading \cite{Ortellado.2025}. The criterion postulates that the crack will kink at an angle $\theta$ such that the stress field in the immediate vicinity of the new crack tip is purely symmetric like locally pure mode I. Mathematically, this local symmetry is achieved when the shear stress intensity factor at the kinked tip vanishes,
\begin{equation}
    K_{II}(\theta)=0.
    \label{eq:pls_condition}
\end{equation}

Similarly, the crack growth angle $\theta_P$ is,
\begin{equation}
    \theta_P = \mathrm{argmin}_{\theta}\, |K_{II}(\theta)|.
    \label{eq:pls_crack_growth_angle}
\end{equation}

This condition is notably equivalent to the prediction derived from the MTS criterion \cite{Paris.1963}. Among the possible solutions for $\theta$ from Eq.~\eqref{eq:pls_condition}, the physically admissible one is chosen, which must yield a tensile opening at the new tip, $K_I(\theta)>0$, rather than crack face interpenetration of $K_I(\theta)<0$.

\subsection{Transfer Learning Strategy}
\label{sec:transfer_learning}

Transfer learning strategy is a powerful and efficient technique in the field of ML \cite{Zhu.2025b}. Its core idea is to use the knowledge from a pre-trained model as a starting point to fine-tune a new model on a new task. Instead of starting from scratch, the new model can leverage the pre-trained model's learned representations and parameters to achieve better and faster training performance on the new task. This allows the new model to benefit from prior knowledge rather than training from scratch.

In the crack propagation loop, as the crack grows with small steps of size $\Delta a$, the geometry and boundary conditions change only slightly from the previous step.  Therefore, it's better to reuse the network weights and SIFs from the previous step as an informed initialization for the next step. This maintains the continuity of the auxiliary field while reducing unnecessary parameter searches.

It should be noted that Cotterell-Rice (C-R) first-order perturbation method \cite{Cotterell.1980} is used to map the SIFs at the new tip of a slightly kinked crack. The equations are expressed as,
\begin{equation}
    \left\{
    \begin{aligned}
        K_I'(\theta)    & = K_I \cos^3\frac{\theta}{2} - 3K_{II}\sin\frac{\theta}{2}\cos^2\frac{\theta}{2},                         \\[6pt]
        K_{II}'(\theta) & = K_I \sin\frac{\theta}{2}\cos^2\frac{\theta}{2} + K_{II}\Bigl(1-\frac{3}{2}\sin^2\frac{\theta}{2}\Bigr).
    \end{aligned}
    \right.
    \label{eq:cr_kink_map}
\end{equation}
where $K_I$ and $K_{II}$ are the SIFs in the previous step, and $K_I'$ and $K_{II}'$ are the SIFs after the kink. The C-R method is necessary during TL process because the SIFs are defined with respect to the local crack frame. A small kink changes the frame. Direct reuse of the old SIFs in the new step leads to mode-mixing errors \cite{Si.2024}. The opening and shear parts no longer match the actual near tip asymptotics \cite{Mitchell.2017}. The C-R mapping transports the amplitudes from the old frame to the new frame with first-order accuracy in the kink angle. The mapped pair $K_I'$ and $K_{II}'$ can match the leading square root singularity at the new tip. This is essential not only for maintaining the continuity of the auxiliary field, but also for ensuring consistent Mode I growth and preventing spurious shear artifacts after each increment.

Thus, the crack propagation loop is defined as follows. First, the geometry and interface is updated. Second, the SIFs are mapped to the new step using the C-R method as $K' = K_I' + iK_{II}'$. Then, the Williams parts and holomorphic branches start already close to a feasible state, ensuring the optimizer needs only a short fine-tuning. Details of computational algorithm for the KMINN derived by TL are provided in Table~\ref{tbl:algo_nested_tl}.

\begin{table*}[ht]
    \caption{Nested-loop crack growth with transfer learning and SIF seeding.}
    \label{tbl:algo_nested_tl}
    \begin{tabularx}{\textwidth}{@{}X@{}}
        \toprule
        \textbf{Algorithm: KMINN crack growth with transfer learning}                                                                                                                                                                                                                      \\
        \midrule
        \textbf{input}: geometry, loads, $E$, $\nu$, $n_{\text{steps}}$, $\Delta a$, schedules $(N_{\mathrm{Adam}},N_{\mathrm{LBFGS}})$, reduced $(\tilde N_{\mathrm{Adam}},\tilde N_{\mathrm{LBFGS}})$, I-integral radius $r$, criterion $\in\{\mathrm{MTS},\mathrm{MERR},\mathrm{PLS}\}$ \\
        \textbf{output}: crack path, $\{\theta_s\}$, $\{K_I^s,K_{II}^s\}$                                                                                                                                                                                                                  \\
        \addlinespace[2pt]
        1\quad \textbf{for} $s=1{:}n_{\text{steps}}$ \textbf{do}                                                                                                                                                                                                                           \\
        2\qquad \textbf{if} $s=1$ \textbf{then}                                                                                                                                                                                                                                            \\
        3\qquad\quad build BCs with Williams \& tips; init weights \& tip SIFs                                                                                                                                                                                                             \\
        4\qquad\quad train with $\{\text{Adam},\text{L-BFGS}\}$                                                                                                                                                                                                                            \\
        5\qquad\quad $(K_I^1,K_{II}^1)\leftarrow$ I-integral @ $r$                                                                                                                                                                                                                         \\
        6\qquad\quad $\theta_1 \leftarrow$ criterion$(K_I^1,K_{II}^1)$                                                                                                                                                                                                                     \\
        7\qquad \textbf{else}                                                                                                                                                                                                                                                              \\
        8\qquad\quad extend crack by $\Delta a$ along $\theta_{s-1}$; update DD spine/overlap                                                                                                                                                                                              \\
        9\qquad\quad map SIFs via Cotterell--Rice using $\theta_{s-1}$                                                                                                                                                                                                                     \\
        10\qquad\quad instantiate new boundary; load previous weights; seed $(K_I',K_{II}')$                                                                                                                                                                                               \\
        11\qquad\quad fine-tune with $\{\text{Adam},\text{L-BFGS}\}$                                                                                                                                                                                                                       \\
        12\qquad\quad $(K_I^s,K_{II}^s)\leftarrow$ I-integral @ $r$;                                                                                                                                                                                                                       \\
        13\qquad\quad $\theta_s \leftarrow$ criterion$(K_I^s,K_{II}^s)$                                                                                                                                                                                                                    \\
        14\qquad \textbf{end if}                                                                                                                                                                                                                                                           \\
        15\qquad \textbf{stop} if tip hits boundary or $s=n_{\text{steps}}$                                                                                                                                                                                                                \\
        16\quad \textbf{end for}                                                                                                                                                                                                                                                           \\
        \bottomrule
    \end{tabularx}
\end{table*}

\section{Case Studies}
\label{sec:case_studies}
In this section, the KMINN with Williams enrichment is utilized to evaluate the SIFs and achieve the simulation of crack propagation. All case studies were conducted on the high-performance computing (HPC) cluster at RWTH Aachen University, utilizing NVIDIA H100 GPUs. The implementation is based on Python 3.13.5 and PyTorch 2.7.0, accelerated with CUDA 12.6 and cuDNN 9.5.1. Since the present study focuses on numerical methodology rather than material calibration, all cases are solved for a homogeneous isotropic linear-elastic solid with Young's modulus $E = 210$ GPa and Poisson ratio $\nu = 0.3$. Additionally, the element size is set to 0.1 mm in all FEM simulations, giving more than 5000 CPE4R elements.

\subsection{Evaluation of SIFs}
\label{sec:sifs_crack_evaluation}

\begin{figure*}[ht]
    \centering
    \includegraphics[width=0.9\textwidth]{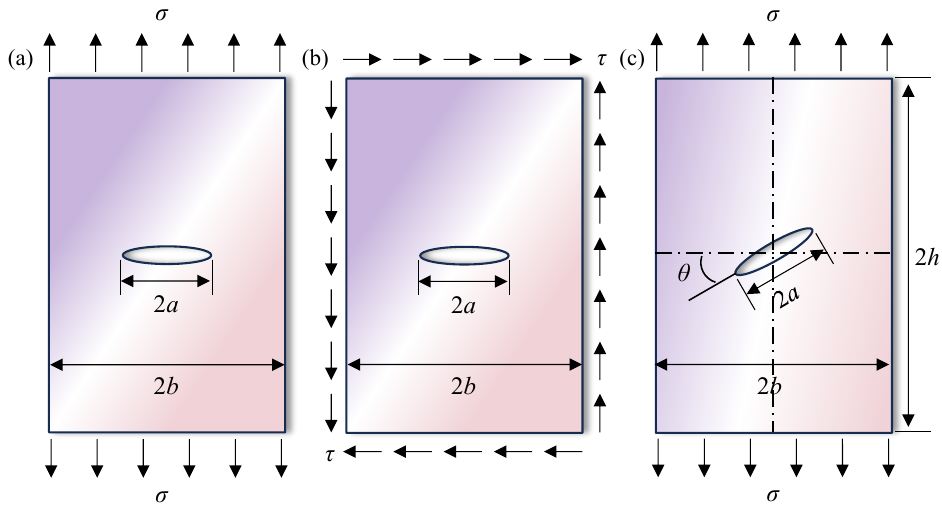}
    \caption{Three typical fracture mechanics case studies for KMINN with Williams enrichment. (a) Center crack tensile (CCT) case, (b) Center crack shear (CCS) case, (c) Oblique center crack tensile (OCCT) case.}
    \label{fig:case_studies}
\end{figure*}

To evaluate the accuracy of the KMINN with Williams enrichment, the SIFs are computed for three benchmark problems: center crack tensile (CCT), center crack shear (CCS), and oblique center crack tensile (OCCT), as illustrated in Fig.~\ref{fig:case_studies}. The considered geometry is a 2D plane-strain rectangular plate with half-width $b = 4~\mathrm{mm}$ and half-height $h = 6~\mathrm{mm}$, containing a centered crack of half-length $a = 1.5~\mathrm{mm}$. A remote load of 1~MPa is applied for all cases. Since the governing PDEs are automatically satisfied by holomorphic functions, only boundary points are used for training and testing. Specifically, 1000 points are sampled for training and 100 for testing. The default KMINN employs three hidden layers with ten neurons each (3~$\times$~10) and a learning rate of $10^{-2}$. A mixed training strategy is adopted as first 2500 epochs with the Adam optimizer followed by 2500 epochs with L-BFGS. In the first CCT case, both the mixed and full-Adam strategies are compared to assess the influence of the optimizer. All results are compared against analytical and FEM references.

\subsubsection{Center Crack Tensile}

The first case is a CCT problem, which is a classical problem in fracture mechanics. The problem is defined as a plate in 2D plane strain condition with a center crack under tensile stress load of 1 MPa. The geometry of the problem is shown in Fig.~\ref{fig:case_studies} (a). The theoretical mode I SIF for a finite-width plate is given by \cite{Tada.1973},
\begin{equation}
    K_{I,~\mathrm{Ana}} = \sigma \sqrt{\pi a}
    \left[ 1 - 0.025 \left( \frac{a}{b} \right)^2 + 0.06 \left( \frac{a}{b} \right)^4 \right]
    \sqrt{ \sec \frac{\pi a}{2b} },
    \label{eq:sif_center_crack_tensile}
\end{equation}
where $\sigma$ is the applied stress, $a$ the half-crack length, and $b$ the half plate width.

\begin{figure}[htbp]
    \centering
    \includegraphics[width=0.45\textwidth]{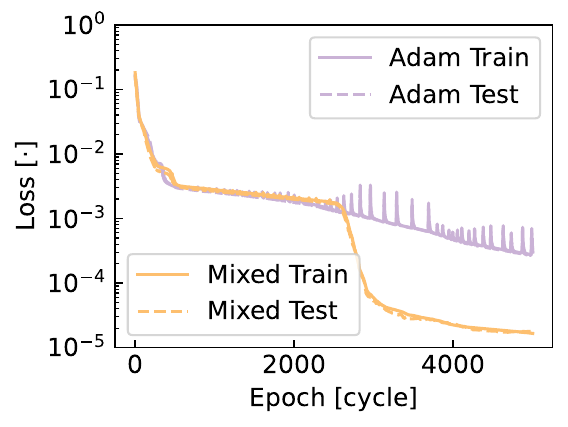}
    \caption{Training loss curves of two training strategies for the center crack tensile problem.}
    \label{fig:training_loss_cct}
\end{figure}

The training and testing loss curves for the two strategies are shown in Fig.~\ref{fig:training_loss_cct}. Both exhibit similar trends, indicating that boundary sampling alone is sufficient to train the model effectively. The Adam-only strategy achieves stable convergence before 2500 epochs but shows fluctuations thereafter, whereas the mixed strategy yields smoother convergence and lower final losses. Therefore, the mixed optimizer is adopted for all subsequent cases.

\begin{figure*}[htbp]
    \centering
    \includegraphics[width=0.99\textwidth]{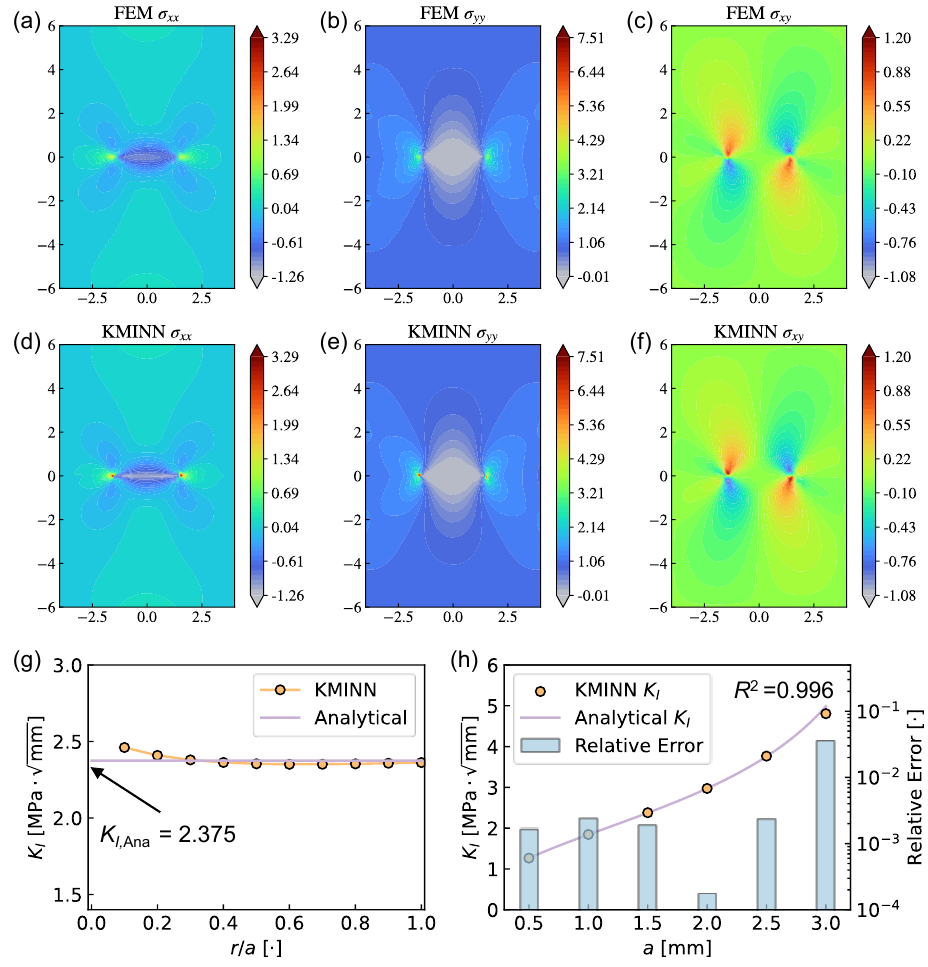}
    \caption{Simulation results for the CCT case. (a)--(c) FEM stress fields, (d)--(f) KMINN stress fields, (g) SIFs comparison with different I-integral radii, (h) SIFs relative error comparison with different crack half-lengths.}
    \label{fig:case_1_figure_combination}
\end{figure*}

To further validate the predictive capability of the KMINN model, the predicted stress fields are compared with FEM results under identical geometry and boundary conditions. As shown in Figs.~\ref{fig:case_1_figure_combination}(a)--(f), the KMINN reproduces the stress singularity near the crack tips and matches the global stress distribution from FEM results closely.

The choice of the integration radius in the $I$-integral method strongly influences the accuracy of SIF evaluation. From Eq.~(\ref{eq:sif_center_crack_tensile}), the theoretical $K_{I,~\mathrm{Ana}} = 2.375~\mathrm{MPa}\cdot\sqrt{\mathrm{mm}}$. Ideally, the computed SIF should be independent of the integration path and radius. However, a too small radius leads to numerical noise due to singularities, whereas a too large radius introduces boundary effects. As shown in Fig.~\ref{fig:case_1_figure_combination}(g), the predicted SIF slightly exceeds the analytical value for $r/a < 0.3$, but good agreement is achieved for $0.4 \le r/a \le 1.0$. Thus, $r/a = 0.4$ is selected as the optimal radius for this case.

Finally, the SIFs for varying half-crack lengths from 0.5 mm to 3.0 mm are compared with analytical predictions in Fig.~\ref{fig:case_1_figure_combination}(h). Excellent agreement is obtained with a global $R^2 = 0.996$. All relative errors are below 1\%, except for $a = 3.0$ mm, where boundary effects slightly increase the deviation to less than 3\%, which is still acceptable. These results demonstrate that the Williams-enriched KMINN accurately captures the stress fields and SIFs for the CCT problem.

\subsubsection{Center Crack Shear}

\begin{figure*}[htbp]
    \centering
    \includegraphics[width=0.99\textwidth]{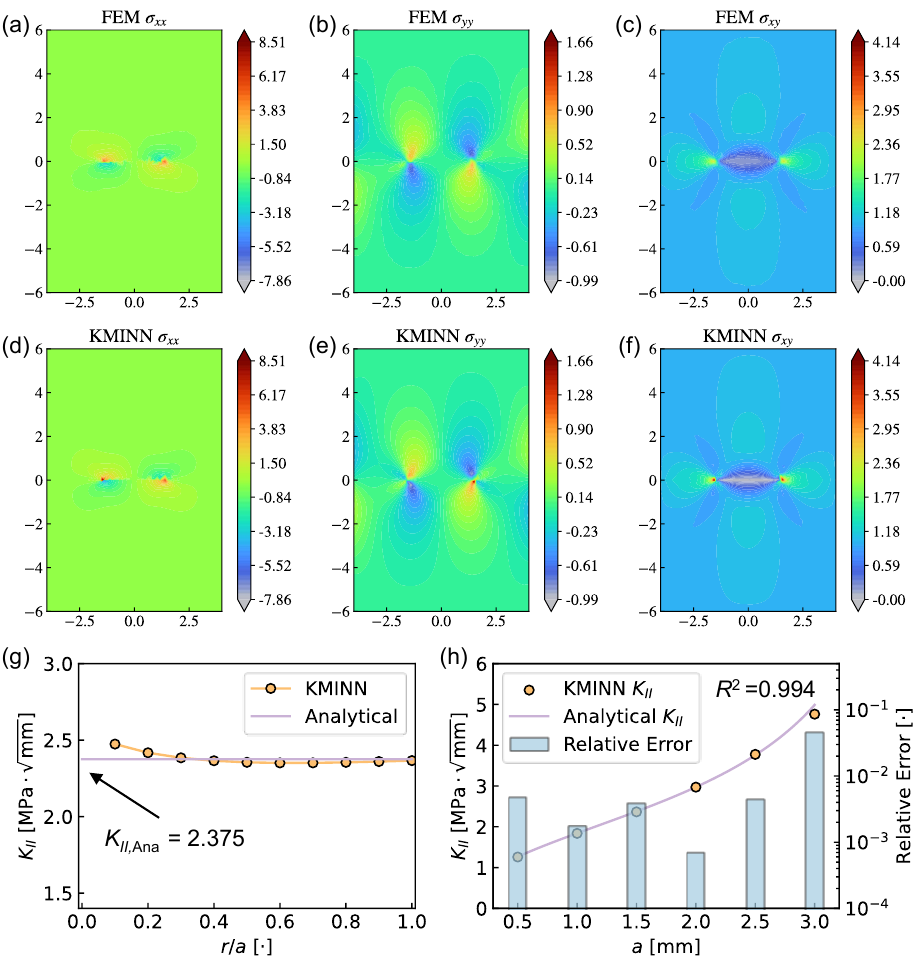}
    \caption{Simulation results for the CCS case. (a)--(c) FEM stress fields, (d)--(f) KMINN stress fields, (g) SIFs comparison with different I-integral radii, (h) SIFs relative error comparison with different crack half-lengths.}
    \label{fig:case_2_figure_combination}
\end{figure*}

The second benchmark case is the CCS problem, which is employed to further evaluate the accuracy of the KMINN with Williams enrichment, as shown in Fig.~\ref{fig:case_studies}(b). The plate geometry is identical to the CCT case, representing a 2D plane-strain rectangular plate with a centered crack of half-length $a = 1.5~\mathrm{mm}$. A remote shear load of $\tau = 1~\mathrm{MPa}$ is applied on the four edges of the plate. Note that it is a clockwise shear load on the top and bottom edges, and a counter-clockwise shear load on the left and right edges to achieve an equilibrium. The theoretical mode II SIF for a finite-width plate is expressed as \cite{Tada.1973},
\begin{equation}
    K_{II,~\mathrm{Ana}} = \tau \sqrt{\pi a}
    \left[ 1 - 0.025 \left( \frac{a}{b} \right)^2 + 0.06 \left( \frac{a}{b} \right)^4 \right]
    \sqrt{ \sec \frac{\pi a}{2b} }.
    \label{eq:sif_center_crack_shear}
\end{equation}
which yields $K_{II,~\mathrm{Ana}} = 2.375~\mathrm{MPa}\cdot\sqrt{\mathrm{mm}}$ for the present geometry. Figs.~\ref{fig:case_2_figure_combination}(a)--(f) compares the predicted stress fields from KMINN and FEM. The results confirm that the KMINN model successfully reproduces the expected stress singularities near the crack tips, while the global stress distribution agrees well with the FEM reference.

To determine the optimal integration radius for the $I$-integral method, the evaluated SIFs at various radii are compared with the analytical value, as shown in Fig.~\ref{fig:case_2_figure_combination}(g). Similar to the CCT case, the results are sensitive to the integration radius when it is very small due to numerical noise near the crack tip. Good agreement is obtained for $0.4 \le r/a \le 1.0$, indicating that $r/a = 0.4$ provides the most accurate and stable SIF evaluation in this case.

Finally, Fig.~\ref{fig:case_2_figure_combination}(h) presents the SIFs obtained for different crack half-lengths ranging from $a = 0.5$ to $3.0~\mathrm{mm}$, compared with the analytical predictions. Excellent agreement is achieved with an overall $R^2 = 0.994$. The relative errors remain below 1\% for all cases except $a = 3.0~\mathrm{mm}$, where the boundary and singularity effects slightly increase the deviation to less than 4\%. These results further verify the capability of the Williams-enriched KMINN to accurately evaluate mode II SIFs under shear loading conditions in CCS problem.

\subsubsection{Oblique Center Crack Tensile}

To evaluate the performance of the proposed KMINN framework under mixed-mode loading, the OCCT problem is analyzed, as shown in Fig.~\ref{fig:case_studies}(c). The geometry corresponds to a 2D plane-strain rectangular plate containing a central crack of half-length $a = 1.5~\mathrm{mm}$, inclined at an angle of $\theta = 30^{\circ}$. A remote tensile stress of $1~\mathrm{MPa}$ is applied in the oblique direction.

\begin{figure*}[htbp]
    \centering
    \includegraphics[width=0.99\textwidth]{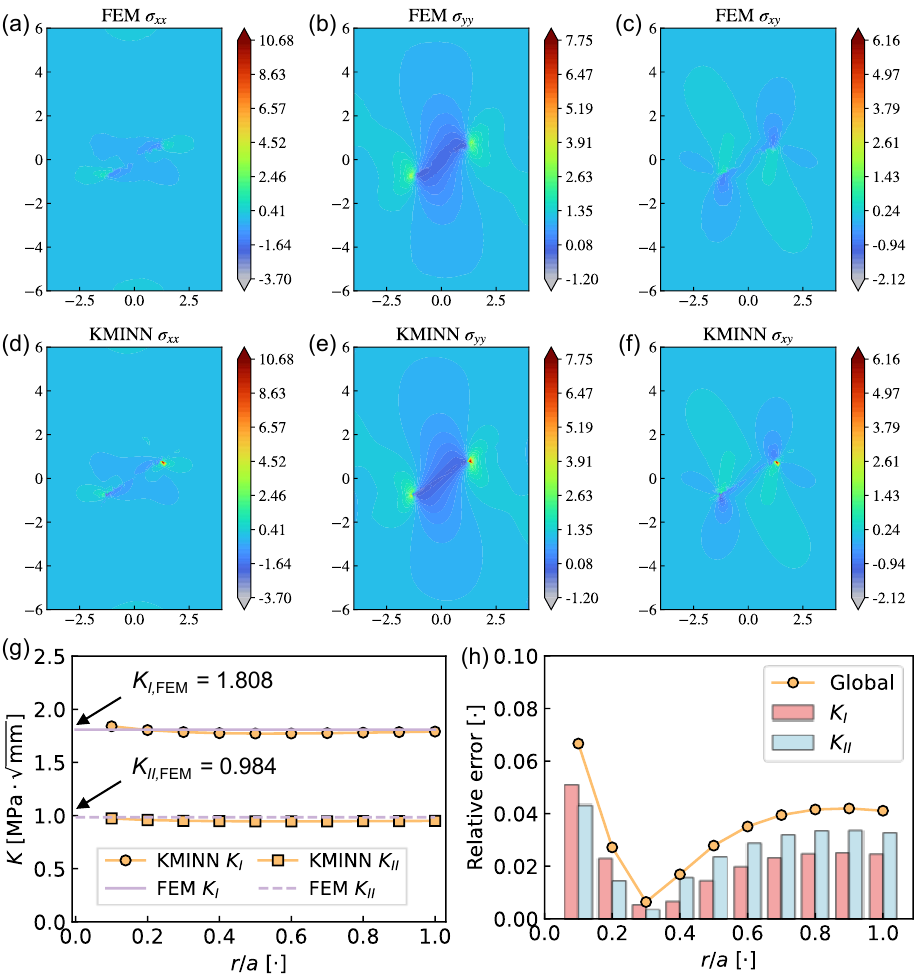}
    \caption{Simulation results for the OCCT case. (a)--(c) FEM stress fields, (d)--(f) KMINN stress fields, (g) SIFs comparison with different I-integral radii, (h) SIFs relative error comparison with different I-integral radii.}
    \label{fig:case_3_figure_combination}
\end{figure*}

Figs.~\ref{fig:case_3_figure_combination}(a)--(f) compares the predicted and FEM stress fields for the OCCT configuration. The results show that the KMINN with Williams enrichment accurately reproduces the stress distribution and singularity near the crack tips, achieving close agreement with the FEM reference results.

The mode I and mode II SIFs obtained from FEM are $K_{I,~\mathrm{FEM}} = 1.808~\mathrm{MPa}\cdot\sqrt{\mathrm{mm}}$ and $K_{II,~\mathrm{FEM}} = 0.984~\mathrm{MPa}\cdot\sqrt{\mathrm{mm}}$, respectively. These are compared with the KMINN-predicted values in Fig.~\ref{fig:case_3_figure_combination}(g). To determine the optimal integration radius for the $I$-integral method, the global relative error is defined as,
\begin{equation}
    \text{err}_{global} = \sqrt{\text{err}_{I}^2 + \text{err}_{II}^2},
    \label{eq:global_relative_error}
\end{equation}
where $\text{err}_{I}$ and $\text{err}_{II}$ denote the relative errors of $K_{I}$ and $K_{II}$, respectively. The variation of $\text{err}_{\mathrm{global}}$ with respect to the integration radius is shown in Fig.~\ref{fig:case_3_figure_combination}(h). The minimum global error occurs at $r/a = 0.3$, which is therefore selected as the optimal radius for the $I$-integral evaluation in this case.

\subsection{Crack Propagation with Transfer Learning}
\label{sec:crack_propagation_direction}

\begin{figure*}[h]
    \centering
    \includegraphics[width=0.9\textwidth]{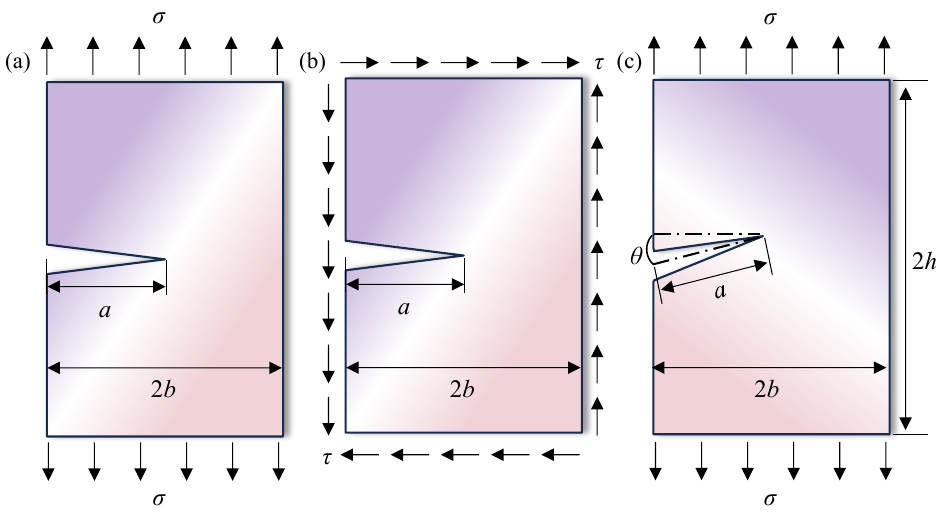}
    \caption{Case studies for KMINN with Williams enrichment. (a) Single edge notch tensile (SENT) case, (b) single edge notch shear (SENS) case, (c) oblique single edge notch tensile (OSENT) case.}
    \label{fig:case_studies_CG}
\end{figure*}

The accuracy of SIF evaluation in the three benchmark cases was verified in Sec.~\ref{sec:sifs_crack_evaluation}. In this section, three fracture criteria are integrated into the KMINN using a TL approach to predict crack propagation directions, thereby enabling subsequent crack growth simulations. To investigate the applicability of KMINN in crack propagation, three representative cases are considered: single-edge notch tensile (SENT), single-edge notch shear (SENS), and oblique single-edge notch tensile (OSENT), corresponding to mode I, mode II, and mixed-mode loading, respectively, as illustrated in Fig.~\ref{fig:case_studies_CG}. The plate has a half-width of 4~mm and a half-height of 6~mm. All cases are subjected to a uniform stress of 1~MPa. Boundary sampling employs 2000 points for training and 200 for testing. The KMINN employs three hidden layers with ten neurons each (3~$\times$~10) and a learning rate of $10^{-2}$. Crack growth is advanced with a fixed increment of $\Delta a = 0.1$~mm over 50 steps. A mixed training strategy is employed to train the KMINN at each step. In the first step, both optimizers run for 2500 epochs; at subsequent steps, the TL method is applied by warm-starting from the previous step's weights and SIFs, followed by fine-tuning for 500 Adam epochs and 500 L-BFGS iterations. The three fracture criteria are incorporated to predict the crack propagation path, and the corresponding results are discussed below.

\subsubsection{Single Edge Notch Tensile}

\begin{figure*}[htbp]
    \centering
    \includegraphics[width=0.99\textwidth]{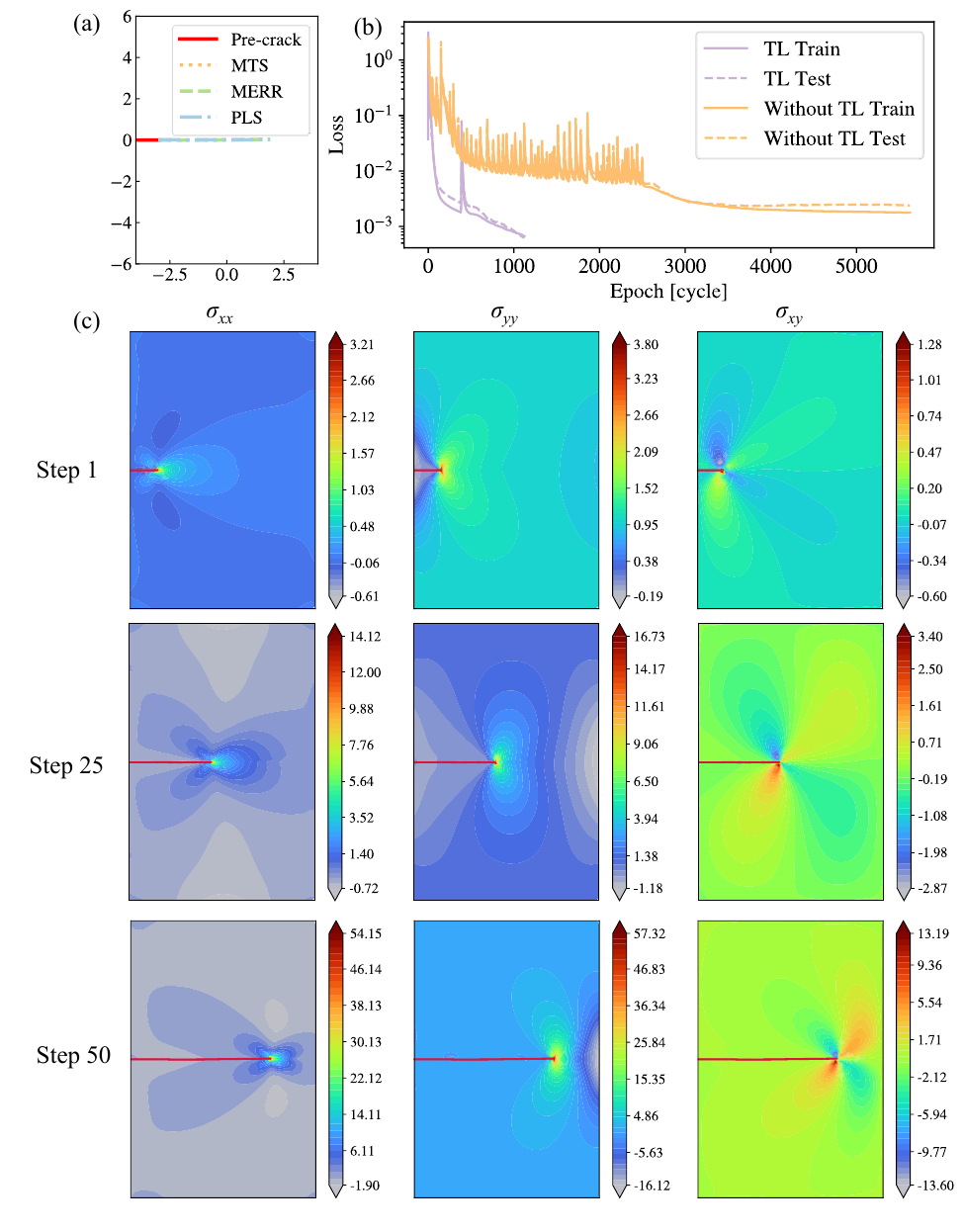}
    \caption{Crack propagation results for the single edge notch tensile (SENT) case. (a) Crack propagation path for the three fracture criteria, (b) loss curves of MTS criterion with or without TL method, (c) stress fields and crack path (red line) of 1st, 25th, and 50th steps for the MTS criterion with TL method.}
    \label{fig:case_4_figure_combination}
\end{figure*}

In the SENT case, the geometry is modeled as a 2D plane strain problem with an initial edge crack $a_0$ = 1~mm and a uniform tensile stress load of 1~MPa.
Fig.~\ref{fig:case_4_figure_combination} overlays the predicted crack paths for MTS/MERR/PLS and reports the loss histories with/without TL and representative stress fields at steps 1/25/50. In Fig.~\ref{fig:case_4_figure_combination}(a), the three crack propagation paths derived from three types of fracture criteria have been successfully predicted by the KMINN model, indicating the KMINN model with Williams enrichment can effectively predict the crack propagation path for the SENT case. Additionally, all three criteria yield nearly coincident crack propagation paths for isotropic SENT, indicating that KMINN with Williams enrichment robustly recovers the path under pure Mode I.

Fig.~\ref{fig:case_4_figure_combination}(b) shows the loss curves of training and testing points about the MTS criterion with or without the TL method at the 50th steps. With the TL method, merely 500 iterations each of Adam and L-BFGS significantly outperform the performance achieved by over 5000 epochs without it, reaching an MSE of $10^{-4}$. The loss curves indicate that the TL method achieves faster convergence compared to the without TL approach. Notably, the TL method also exhibits more stable convergence, even when in the Adam optimizer stage. This illustrates that the TL method can effectively reduce training time and enhance training efficiency. Fig.~\ref{fig:case_4_figure_combination}(c) shows the representative stress fields of the SENT case with MTS criterion and TL method at 1st, 25th, and 50th steps. The crack path is illustrated as a red line in each subfigure. The stress fields exhibit the expected near-tip singularity and global symmetry consistent with the loading, demonstrating the effectiveness of our approach in simulating the entire cracking process.

\subsubsection{Single Edge Notch Shear}

\begin{figure*}[htbp]
    \centering
    \includegraphics[width=0.99\textwidth]{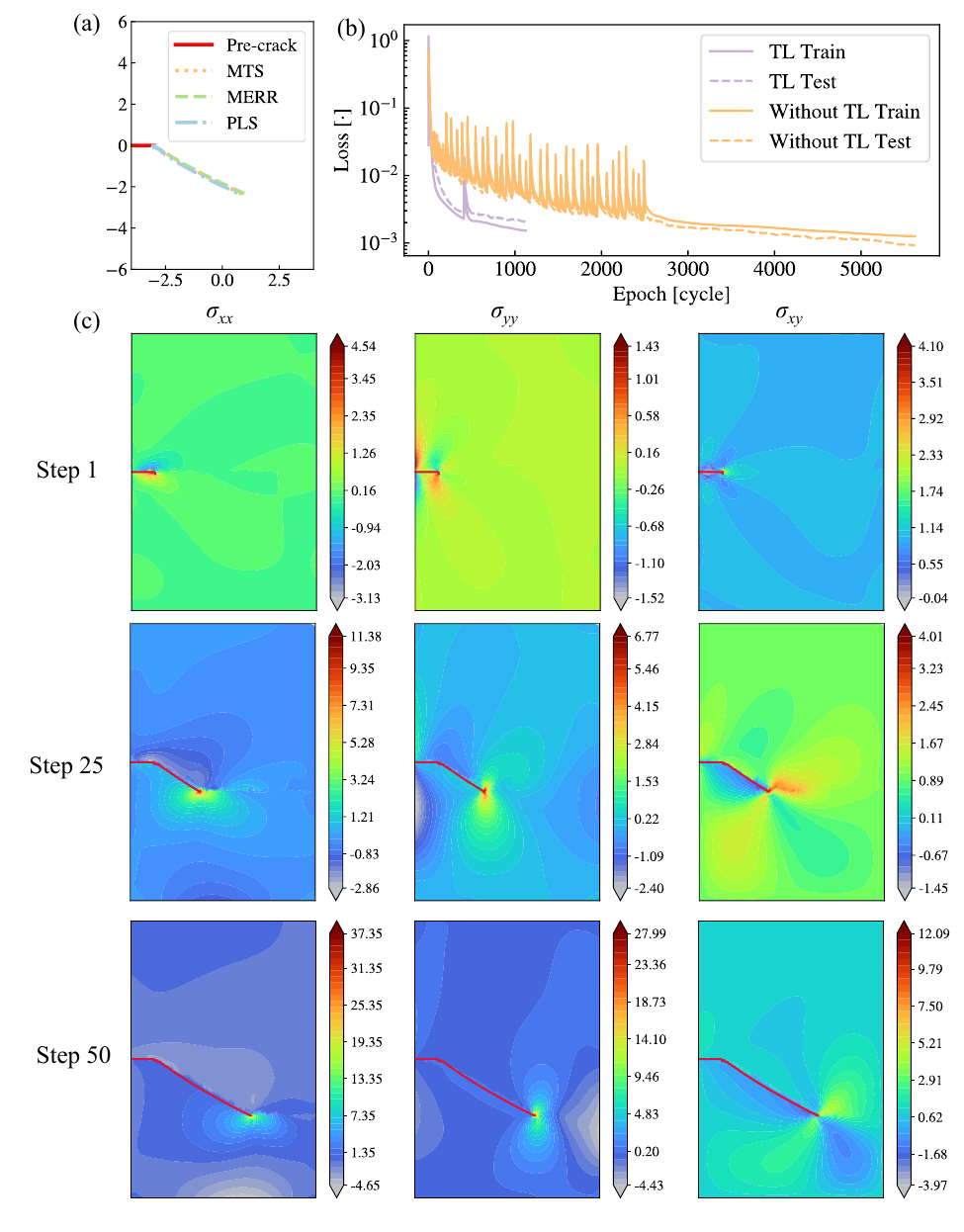}
    \caption{Crack propagation results for the single edge notch shear (SENS) case. (a) Crack propagation path for the three fracture criteria, (b) loss curves of MERR criterion with or without TL method, (c) stress fields and crack path (red line) of 1st, 25th, and 50th steps for the MERR criterion with TL method.}
    \label{fig:case_5_figure_combination}
\end{figure*}

The initial edge crack is $a_0 = 1$ mm and a remote shear traction of $1$ MPa is applied for the single edge notch shear (SENS) case.
Fig.~\ref{fig:case_5_figure_combination} illustrates the crack propagation path, loss curves, and stress fields at steps 1/25/50 for the SENS case. The crack path exhibits a pronounced deflection from the initial crack path in Fig.~\ref{fig:case_5_figure_combination}(a). Furthermore, the three criteria predict modest divergence path only at beginning steps, whereas the full propagation path is still similar to each other.

At step~50 of MERR criterion, Fig.~\ref{fig:case_5_figure_combination}(b) confirms that TL accelerates and stabilizes the convergence relative to training from scratch. Stress fields at each step of MERR criterion in Fig.~\ref{fig:case_5_figure_combination}(c) display the expected anti-symmetric patterns of $\sigma_{xy}$ near the tip and a loss of global symmetry after propagation. Furthermore, $\sigma_{xx}$ and $\sigma_{yy}$ redistribute accordingly with the evolving crack path geometry, consistent with Mode~II-dominated growth.

\subsubsection{Oblique Single Edge Notch Tensile}
\begin{figure*}[htbp]
    \centering
    \includegraphics[width=0.99\textwidth]{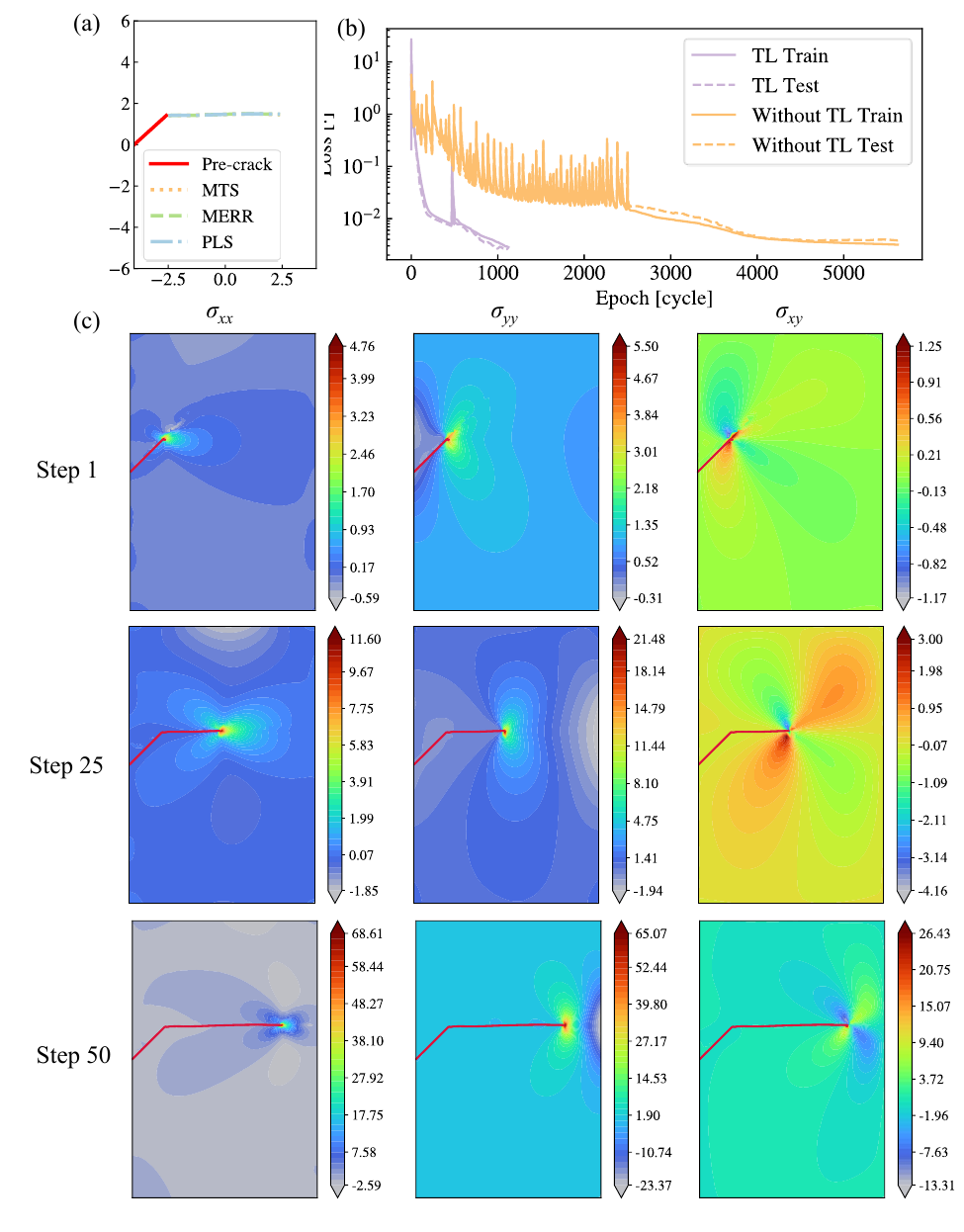}
    \caption{Crack propagation results for the oblique single edge notch tensile (OSENT) case. (a) Crack propagation path for the three fracture criteria, (b) loss curves of PLS criterion with or without TL method, (c) stress fields and crack path (red line) of 1st, 25th, and 50th steps for the PLS criterion with TL method.}
    \label{fig:case_6_figure_combination}
\end{figure*}

An oblique edge crack of length $a_0 = 2$ mm is oriented at $\theta_0 = 45^\circ$ under remote tension $1$ MPa for the oblique single edge notch tensile (OSENT) case. The crack propagation results of OSENT case are shown in Fig.~\ref{fig:case_6_figure_combination}.
As shown in Fig.~\ref{fig:case_6_figure_combination}(a), the crack paths initially display a small deflection from their geometric orientation, a behavior caused by the mixed-mode loading. Following this, the crack path systematically deflect away from their initial alignment and curve towards the loading direction, eventually propagating perpendicular to the applied load. This observed behavior is consistent with our previous experimental results on mixed-mode fracture propagation \cite{Wang.2024j,Wang.2024l}.

Moreover, Fig.~\ref{fig:case_6_figure_combination}(b) again shows faster convergence with TL at step~50 of PLS criterion. The stress field evolution of PLS criterion in Fig.~\ref{fig:case_6_figure_combination}(c) illustrate mixed-mode features. $\sigma_{yy}$ retains a tensile-dominant character, while $\sigma_{xy}$ carries the obliquity-induced shear. Additionally, after crack propagation, the solution lacks global symmetry, in line with the oblique geometry and mixed loading.

\section{Discussion}
\label{sec:discussion}
The numerical results presented in Sections~\ref{sec:sifs_crack_evaluation} and~\ref{sec:crack_propagation_direction} systematically demonstrate the accuracy and robustness of the proposed KMINN framework across a series of case studies. The model successfully reproduces both the global and local features of the stress fields, achieving excellent agreement with the analytical SIFs and FEM stress-field references for the three static cases. In most cases, the SIFs evaluation has a consistently average relative error that remains below 1\%, and the $R^2$ values for CCT and CCS cases are more than 0.99, confirming that the Williams-enriched formulation of KMINN effectively captures the near-tip singular behavior.

\begin{figure*}[htbp]
    \centering
    \includegraphics[width=1.0\textwidth]{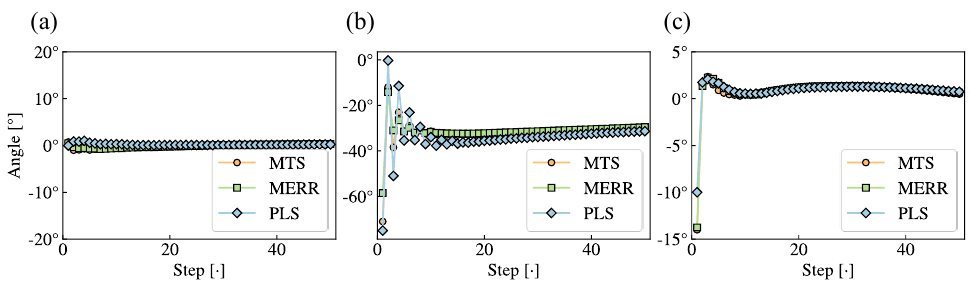}
    \caption{Comparison of the crack propagation angles with different steps for the three fracture criteria for each case, (a) SENT case, (b) SENS case, (c) OSENT case.}
    \label{fig:SIF_evaluation_comparison}
\end{figure*}

Additionally, the crack propagation simulations of the three fracture criteria (MTS, MERR, and PLS) are implemented with the Williams-enriched KMINN framework. The Fig. \ref{fig:SIF_evaluation_comparison} shows the further comparison of the crack propagation angles for the three fracture criteria with different steps for each case. It can be seen that the PLS criterion shows slight fluctuations in the initial stage, especially for the SENS case. However, the MTS and MERR criteria are more stable in this stage. Nevertheless, the crack propagation angles are nearly identical for the three criteria in the stable stage, which is consistent with the results in other literature \cite{Sakha.2022,Aliha.2016,Chen.2022b}. The fundamental reason lies in the symmetry of the crack-tip stress field in isotropic material \cite{Hua.2023}. Regardless of whether the fracture process is interpreted from an energy-based or stress-based perspective, the crack tends to propagate along a path that minimizes the mode II component, approaching zero, thereby extending in a purely opening mode. Cotterell and Rice \cite{Cotterell.1980} demonstrated that, for isotropic and homogeneous materials under mixed-mode loading, the crack deflects toward the direction where $K_{II}=0$. Under this condition, the predictions given by the different fracture criteria coincide. Consequently, for typical brittle materials subjected to pure mode I loading, all criteria predict straight crack growth, whereas under mixed-mode I/II conditions, their predicted deflection angles are nearly identical. In the KMINN simulations, the predicted stress fields near the crack tip show that the shear component quickly decreases to nearly zero along the crack propagation direction. Therefore, the learned solution naturally satisfies the $K_{II}=0$ condition without any additional constraint. That is why the MTS, MERR, and PLS criteria yield almost identical propagation paths in isotropic materials.

From a mechanistic perspective, the model generation is derived from the integration of physically-informed enrichment and knowledge reuse. The Williams enrichment embeds the square-root singularity directly into the neural network representation, thereby reducing the learning task on the network and ensuring that the crack-tip asymptotics are satisfied by construction. This eliminates the requirement for refined sampling near the crack tips and stabilizes the SIFs extraction using the $I$-integral method. On the other hand, the TL method reuses previously trained network weights and SIF information from earlier crack-growth steps, allowing the model to retain the stress-field memory of prior configurations. As clearly shown in Figs.~\ref{fig:case_4_figure_combination}--\ref{fig:case_6_figure_combination}(b), the TL method significantly accelerates the convergence, requiring only about $1/5$ of the number compared to the approach without TL. Table \ref{tb:training_time} details the simulation times for crack propagation cases under different criteria, comparing scenarios with and without TL. Notably, adopting the TL strategy reduced computation time by more than 70\%. This acceleration is due to the physical similarity between subsequent crack growth configurations. Since the stress and displacement fields evolve smoothly with respect to the crack increment, the pretrained weights already approximate the correct functional subspace. As a result, the TL method effectively acts as a low-dimensional projection operator in the parameter space, confining the subsequent optimization to a physically consistent manifold \cite{Zong.2026}. Furthermore, the TL method also exhibits more stable and smooth loss curves, making the training process more efficient and reliable.

\begin{table}[htbp]
    \centering
    \small
    \caption{The simulation time of crack propagation cases under different criteria with or without TL. (units: minutes)}
    \label{tb:training_time}
    \begin{tabular}{lccccccccc}
        \toprule
                   & \multicolumn{3}{c}{SENT} & \multicolumn{3}{c}{SENS} & \multicolumn{3}{c}{OSENT}                                                 \\
        \cmidrule(lr){2-4} \cmidrule(lr){5-7} \cmidrule(lr){8-10}
                   & MTS                      & MERR                     & PLS                       & MTS   & MERR  & PLS   & MTS   & MERR  & PLS   \\ \midrule
        With TL    & 20.1                     & 20.1                     & 20.0                      & 27.4  & 33.5  & 33.8  & 34.5  & 33.1  & 33.5  \\
        Without TL & 87.0                     & 86.3                     & 89.9                      & 138.4 & 144.2 & 135.0 & 114.9 & 138.5 & 139.2 \\ \bottomrule
    \end{tabular}

\end{table}

Compared with conventional FEM and real-valued PINNs, the proposed KMINN framework offers distinct advantages. First, it satisfies the governing PDEs by construction through the KM potential formulation, reducing computational cost by eliminating volume sampling. The refined sampling scheme is unnecessary in the KMINN framework, particularly for the crack propagation problems, compared to the FEM and real-valued PINN methods. The latter rely on the carefully designed sampling schemes to ensure the accuracy of the solution \cite{Gu.2024,Shen.2022}. Second, the enriched representation requires a much smaller network to achieve comparable accuracy to FEM meshes with thousands of elements. For instance, in the CCT case, the KMINN framework requires only 1000 training points to achieve comparable accuracy to FEM meshes with more than 5000 elements. Third, the TL-based propagation strategy provides a unified approach for sequential crack-growth problems without re-meshing or re-training from scratch, which is typically unavoidable in FEM \cite{Wang.2024m}. These characteristics make the KMINN framework not only computationally efficient but also inherently scalable to mixed-mode fracture problems.

Nevertheless, some limitations remain. The present study is restricted to 2D plane-strain, linear-elastic fracture problems for isotropic materials. How to extend the KMINN framework to plasticity and anisotropic problems is still a challenging problem. Moreover, the first-order Williams enrichment without T-stress term is only used in current work. T-stress is a constant stress term parallel to the crack surface. It does not have a singularity but has an important influence on the stress state at the crack tip.  James et al. \cite{James.2013} introduced the shielding effects of the plastic region on stress fields in various directions by using the $\ln r$ term in the enrichment term. Thus, the KMINN framework with the plastic model or higher-order Williams enrichment is a potential research direction by adding plasticity-related and T-stress enrichment terms. Despite these challenges, the results highlight that KMINN provides a promising foundation for next-generation PINN frameworks for fracture mechanics, with potential applications in damage tolerance analysis and fatigue life prediction.

\section{Conclusions}
\label{sec:conclusions}

This study built a KMINN framework with Williams enrichment and the TL method for accurate and efficient fracture analysis. The developed model reproduces analytical and FEM reference results across a series of benchmark cases, achieving average relative errors below 1\% in SIF evaluation and $R^2>0.99$ for both mode I and mode II loadings. The enrichment strategy ensures that the near-tip singularity is captured by construction, while the TL scheme reduces the required training time by more than 70\% compared with training from scratch. The method successfully predicts crack propagation paths in SENT, SENS, and OSENT configurations, where the propagation angles obtained from MTS, MERR, and PLS criteria are nearly identical in the stable stage, consistent with theoretical expectations for isotropic materials. These results confirm that KMINN provides an accurate and physically consistent framework for simulating mixed-mode crack growth, offering a model that is PDE-independent in the training process relative to conventional PINN approaches. Future work will focus on extending the formulation to include plasticity-related or $T$-stress enrichment, to further address plasticity and anisotropy effects in complex fracture systems.

\section*{Acknowledgments}
The simulations were performed with computing resources granted by RWTH Aachen University, Germany, under project (rwth1654). The authors sincerely acknowledge the support from the National Natural Science Foundation of China (Grant No. 52375159). The first author sincerely thanks the financial support of the China Scholarship Council (CSC: 202307000038).

\section*{Author Contributions}
Shuwei Zhou: Writing -- original draft, Conceptualization, Formal analysis, Visualization. Christian Häffner: Writing -- review \& editing, Formal analysis. Shuancheng Wang: Writing -- review \& editing, Conceptualization. Sophie Stebner: Writing -- review \& editing, Formal analysis. Zhen Liao: Writing -- review \& editing. Bing Yang: Writing -- review \& editing, Data collection. Zhichao Wei: Writing -- review \& editing, Formal analysis. Sebastian Münstermann: Writing -- review \& editing, Supervision.

\section*{Data Availability}
The data that support the findings of this study are available from the corresponding author upon reasonable request.

\section*{Declaration of Competing Interest}
The authors declare that they have no known competing financial interests or personal relationships that could have appeared to influence the work reported in this paper.
\clearpage
\appendix

\section{Holomorphicity of the Williams Enrichment on Subdomains}
\label{sec:Holomorphicity_proof}
\textbf{Proposition.}
Let $\Omega_j\subset\mathbb{C}$ be a subdomain obtained by removing, for each crack tip $p^{\pm}_{l,j}$ with orientation $a^{\pm}_{l,j}$, the ray,
\begin{equation}
    \mathcal{R}^{\pm}_{j}=\bigl\{\,p^{\pm}_{l,j}+e^{i a^{\pm}_{l,j}}\,t \;:\; t\le 0\,\bigr\},
\end{equation}
and the tip itself. Define the affine map,
\begin{equation}
    h^{\pm}_j(z)=e^{-i a^{\pm}_{l,j}}\bigl(z-p^{\pm}_{l,j}\bigr),\qquad z\in\Omega_j,
\end{equation}
and choose the branch of $\sqrt{\cdot}$ that is holomorphic on $\mathbb{C}\setminus(-\infty,0]$, with the branch cut aligned with $\mathcal{R}^{\pm}_{j}$ so that $h^{\pm}_j(\Omega_j)\cap(-\infty,0]=\varnothing$. Then the Williams enrichment is,
\begin{equation}
    \left\{
    \begin{aligned}
        \varphi^{\pm}_{W,j}(z) & = \dfrac{K^{\pm}_{l,j}}{\sqrt{2\pi}}\;e^{i a^{\pm}_{l,j}}
        \sqrt{\,h^{\pm}_j(z)\,},                                                           \\[6pt]
        \psi^{\pm}_{W,j}(z)    & = \dfrac{1}{\sqrt{2\pi}}\left[
        \Bigl(\bar K^{\pm}_{l,j}-\dfrac{K^{\pm}_{l,j}}{2}\Bigr)\!
        e^{-i a^{\pm}_{l,j}}\sqrt{\,h^{\pm}_j(z)\,}
        - \dfrac{K^{\pm}_{l,j}\,p^{\pm}_{l,j}}{2}\;\dfrac{1}{\sqrt{\,h^{\pm}_j(z)\,}}
        \right],
    \end{aligned}
    \right.
\end{equation}
which is holomorphic on $\Omega_j$, i.e., $\partial_{\bar z}\varphi^{\pm}_{W,j}=\partial_{\bar z}\psi^{\pm}_{W,j}=0$ in $\Omega_j$.

\textbf{Proof.}
Since $h^{\pm}_j$ is complex affine,
\begin{equation}
    \partial_{\bar z} h^{\pm}_j(z)=0,\qquad h^{\pm}_j(z)\neq 0\ \text{on}\ \Omega_j.
\end{equation}
Let $S(w)=\sqrt{w}$ on $\mathbb{C}\setminus(-\infty,0]$ and $R(w)=1/\sqrt{w}$ on the same slit domain. Then,
\begin{equation}
    S'(w)=\dfrac{1}{2\sqrt{w}},\qquad R'(w)=-\dfrac{1}{2}\,w^{-3/2}.
\end{equation}
By the Wirtinger chain rule,
\begin{equation}
    \partial_{\bar z}\!\left[S\!\bigl(h^{\pm}_j(z)\bigr)\right]
    = S'\!\bigl(h^{\pm}_j(z)\bigr)\,\partial_{\bar z} h^{\pm}_j(z)
    = \dfrac{1}{2\sqrt{h^{\pm}_j(z)}}\cdot 0 = 0,
\end{equation}
\begin{equation}
    \partial_{\bar z}\!\left[R\!\bigl(h^{\pm}_j(z)\bigr)\right]
    = R'\!\bigl(h^{\pm}_j(z)\bigr)\,\partial_{\bar z} h^{\pm}_j(z)
    = \left(-\dfrac{1}{2}\,[h^{\pm}_j(z)]^{-3/2}\right)\cdot 0 = 0.
\end{equation}
Multiplying by constants and adding preserves holomorphicity. Hence,
\begin{equation}
    \partial_{\bar z}\varphi^{\pm}_{W,j}(z)=0,\qquad
    \partial_{\bar z}\psi^{\pm}_{W,j}(z)=0,\qquad z\in\Omega_j.
\end{equation}
Therefore, $\varphi^{\pm}_{W,j}$ and $\psi^{\pm}_{W,j}$ are holomorphic in the domain excluding the branch cut and the tip. Thus,
\begin{equation}
    \varphi_j=\varphi_{n,j}+\varphi^{+}_{W,j}+\varphi^{-}_{W,j},\qquad
    \psi_j=\psi_{n,j}+\psi^{+}_{W,j}+\psi^{-}_{W,j}
\end{equation}
are holomorphic.





\begin{thebibliography}{10}
    \expandafter\ifx\csname url\endcsname\relax
      \def\url#1{\texttt{#1}}\fi
    \expandafter\ifx\csname urlprefix\endcsname\relax\def\urlprefix{URL }\fi
    \expandafter\ifx\csname href\endcsname\relax
      \def\href#1#2{#2} \def\path#1{#1}\fi
    
    \bibitem{Wei.2024}
    Z.~Wei, S.~Gerke, M.~Br{\"u}nig, Ductile damage and fracture characterizations in bi-cyclic biaxial experiments, International Journal of Mechanical Sciences 276 (2024) 109380.
    \newblock \href {https://doi.org/10.1016/j.ijmecsci.2024.109380} {\path{doi:10.1016/j.ijmecsci.2024.109380}}.
    
    \bibitem{Wei.2022}
    Z.~Wei, M.~Zistl, S.~Gerke, M.~Br{\"u}nig, Analysis of ductile damage and fracture under reverse loading, International Journal of Mechanical Sciences 228 (2022) 107476.
    \newblock \href {https://doi.org/10.1016/j.ijmecsci.2022.107476} {\path{doi:10.1016/j.ijmecsci.2022.107476}}.
    
    \bibitem{Zhou.2026}
    S.~Zhou, M.~Huang, C.~H{\"a}ffner, S.~Stebner, M.~Cai, Z.~Wei, B.~Yang, S.~M{\"u}nstermann, Microstructure-sensitive crystal plasticity and fatigue indicator modeling for lz50 steel, International Journal of Fatigue 203 (2026) 109302.
    \newblock \href {https://doi.org/10.1016/j.ijfatigue.2025.109302} {\path{doi:10.1016/j.ijfatigue.2025.109302}}.
    
    \bibitem{Wei.2025}
    Z.~Wei, G.~Mao, S.~Gerke, S.~M{\"u}nstermann, M.~Br{\"u}nig, Experimental analysis and modeling of anisotropic ductile damage in non-proportional extreme low-cycle biaxial loading with shear-tension histories, International Journal of Plasticity 194 (2025) 104474.
    \newblock \href {https://doi.org/10.1016/j.ijplas.2025.104474} {\path{doi:10.1016/j.ijplas.2025.104474}}.
    
    \bibitem{Ge.2024}
    X.~Ge, L.~Zhou, Y.~Ying, S.~Bagherifard, M.~Guagliano, Combining phase field method and critical distance theory for predicting fatigue life of notched specimens, International Journal of Mechanical Sciences 282 (2024) 109608.
    \newblock \href {https://doi.org/10.1016/j.ijmecsci.2024.109608} {\path{doi:10.1016/j.ijmecsci.2024.109608}}.
    
    \bibitem{Lo.2019}
    Y.-S. Lo, M.~J. Borden, K.~Ravi-Chandar, C.~M. Landis, A phase-field model for fatigue crack growth, Journal of the Mechanics and Physics of Solids 132 (2019) 103684.
    \newblock \href {https://doi.org/10.1016/j.jmps.2019.103684} {\path{doi:10.1016/j.jmps.2019.103684}}.
    
    \bibitem{Fathi.2025}
    F.~Fathi, R.~d. Borst, G.~Torelli, A consistent phase-field-regularised partition of unity method for fracture analysis, Computer Methods in Applied Mechanics and Engineering 446 (2025) 118267.
    \newblock \href {https://doi.org/10.1016/j.cma.2025.118267} {\path{doi:10.1016/j.cma.2025.118267}}.
    
    \bibitem{Liu.2025}
    J.~Liu, Z.~Gao, S.~Liang, Y.~Zhu, L.~Zhao, M.~Huang, Z.~Li, A hybrid peridynamic framework incorporating entanglement effect for hyperelastic materials, International Journal of Mechanical Sciences 308 (2025) 110955.
    \newblock \href {https://doi.org/10.1016/j.ijmecsci.2025.110955} {\path{doi:10.1016/j.ijmecsci.2025.110955}}.
    
    \bibitem{Xin.2025}
    Y.~Xin, Z.~Li, D.~Huang, Y.~Lu, An electromechanical coupling peridynamic model for piezoelectric solids with defects, International Journal of Mechanical Sciences 304 (2025) 110679.
    \newblock \href {https://doi.org/10.1016/j.ijmecsci.2025.110679} {\path{doi:10.1016/j.ijmecsci.2025.110679}}.
    
    \bibitem{Duan.2025}
    Y.~Duan, C.~Wang, B.~Yin, K.~M. Liew, Peridynamic modeling of interfacial failure in 3d-printed concrete, International Journal of Mechanical Sciences 301 (2025) 110490.
    \newblock \href {https://doi.org/10.1016/j.ijmecsci.2025.110490} {\path{doi:10.1016/j.ijmecsci.2025.110490}}.
    
    \bibitem{Spada.2025}
    A.~Spada, M.~Puccia, E.~Sacco, G.~Giambanco, A coupled fem-vem approach for crack tracking in quasi-brittle materials, Computer Methods in Applied Mechanics and Engineering 437 (2025) 117756.
    \newblock \href {https://doi.org/10.1016/j.cma.2025.117756} {\path{doi:10.1016/j.cma.2025.117756}}.
    
    \bibitem{Yang.2021b}
    B.~Yang, Z.~Wei, F.~A. D{\'i}az, Z.~Liao, M.~N. James, New algorithm for optimised fitting of dic data to crack tip plastic zone using the cjp model, Theoretical and Applied Fracture Mechanics 113 (2021) 102950.
    \newblock \href {https://doi.org/10.1016/j.tafmec.2021.102950} {\path{doi:10.1016/j.tafmec.2021.102950}}.
    
    \bibitem{Wang.2024g}
    S.~Wang, B.~Yang, S.~Zhou, Y.~Wang, S.~Xiao, Effect of stress ratio and overload on mixed-mode crack propagation behaviour of ea4t steel, Engineering Fracture Mechanics 306 (2024) 110210.
    \newblock \href {https://doi.org/10.1016/j.engfracmech.2024.110210} {\path{doi:10.1016/j.engfracmech.2024.110210}}.
    
    \bibitem{Mao.2022}
    J.~Mao, Y.~Xu, D.~Hu, X.~Liu, J.~Pan, H.~Sun, R.~Wang, Microstructurally short crack growth simulation combining crystal plasticity with extended finite element method, Engineering Fracture Mechanics 275 (2022) 108786.
    \newblock \href {https://doi.org/10.1016/j.engfracmech.2022.108786} {\path{doi:10.1016/j.engfracmech.2022.108786}}.
    
    \bibitem{Mirzaei.2025}
    A.~M. Mirzaei, Stress, strain, or displacement? a novel machine learning based framework to predict mixed mode i/ii fracture load and initiation angle, Engineering Fracture Mechanics 325 (2025) 111349.
    \newblock \href {https://doi.org/10.1016/j.engfracmech.2025.111349} {\path{doi:10.1016/j.engfracmech.2025.111349}}.
    
    \bibitem{Zhou.2024c}
    S.~Zhou, B.~Yang, S.~Xiao, G.~Yang, T.~Zhu, Interpretable machine learning method for modelling fatigue short crack growth behaviour, Metals and Materials International 30~(7) (2024) 1944--1964.
    \newblock \href {https://doi.org/10.1007/s12540-024-01628-6} {\path{doi:10.1007/s12540-024-01628-6}}.
    
    \bibitem{Pagan.2022}
    D.~C. Pagan, C.~R. Pash, A.~R. Benson, M.~P. Kasemer, Graph neural network modeling of grain-scale anisotropic elastic behavior using simulated and measured microscale data, npj Computational Materials 8~(259) (2022).
    \newblock \href {https://doi.org/10.1038/s41524-022-00952-y} {\path{doi:10.1038/s41524-022-00952-y}}.
    
    \bibitem{Sim.2025}
    G.-J. Sim, M.-G. Lee, M.~I. Latypov, Fip-gnn: Graph neural networks for scalable prediction of grain-level fatigue indicator parameters, Scripta Materialia 255 (2025) 116407.
    \newblock \href {https://doi.org/10.1016/j.scriptamat.2024.116407} {\path{doi:10.1016/j.scriptamat.2024.116407}}.
    
    \bibitem{Peng.2022}
    X.~Peng, S.~Wu, W.~Qian, J.~Bao, Y.~Hu, Z.~Zhan, G.~Guo, P.~J. Withers, The potency of defects on fatigue of additively manufactured metals, International Journal of Mechanical Sciences 221 (2022) 107185.
    \newblock \href {https://doi.org/10.1016/j.ijmecsci.2022.107185} {\path{doi:10.1016/j.ijmecsci.2022.107185}}.
    
    \bibitem{Liu.2023e}
    Y.~Liu, J.~Fan, G.~Zhu, M.~Zhu, F.~Xuan, Data-driven approach to very high cycle fatigue life prediction, Engineering Fracture Mechanics 292 (2023) 109630.
    \newblock \href {https://doi.org/10.1016/j.engfracmech.2023.109630} {\path{doi:10.1016/j.engfracmech.2023.109630}}.
    
    \bibitem{Henrich.2020}
    M.~Henrich, F.~P{\"u}tz, S.~M{\"u}nstermann, A novel approach to discrete representative volume element automation and generation-dragen, Materials (Basel, Switzerland) 13~(8) (2020).
    \newblock \href {https://doi.org/10.3390/ma13081887} {\path{doi:10.3390/ma13081887}}.
    
    \bibitem{Liang.2023}
    Z.~Liang, X.~Wang, Y.~Cui, W.~Xu, Y.~Zhang, Y.~He, A new data-driven probabilistic fatigue life prediction framework informed by experiments and multiscale simulation, International Journal of Fatigue 174 (2023) 107731.
    \newblock \href {https://doi.org/10.1016/j.ijfatigue.2023.107731} {\path{doi:10.1016/j.ijfatigue.2023.107731}}.
    
    \bibitem{Zhou.2023b}
    S.~Zhou, B.~Yang, S.~Xiao, G.~Yang, T.~Zhu, Crack growth rate model derived from domain knowledge-guided symbolic regression, Chinese Journal of Mechanical Engineering 36~(1) (2023).
    \newblock \href {https://doi.org/10.1186/s10033-023-00876-8} {\path{doi:10.1186/s10033-023-00876-8}}.
    
    \bibitem{Wang.2024j}
    S.~Wang, S.~Zhou, B.~Yang, S.~Xiao, G.~Yang, T.~Zhu, Effective stress intensity factor range for fatigue cracks propagating in mixed mode i-ii loading, Engineering Fracture Mechanics 312 (2024) 110641.
    \newblock \href {https://doi.org/10.1016/j.engfracmech.2024.110641} {\path{doi:10.1016/j.engfracmech.2024.110641}}.
    
    \bibitem{Fehlemann.2025}
    N.~C. Fehlemann, I.~Biermann, S.~M{\"u}nstermann, Exploring structure--property relations in dual phase steels using crystal plasticity and variance based global sensitivity analysis, Materials {\&} Design 259 (2025) 114794.
    \newblock \href {https://doi.org/10.1016/j.matdes.2025.114794} {\path{doi:10.1016/j.matdes.2025.114794}}.
    
    \bibitem{Kong.2025}
    L.~Kong, B.~Pan, M.~Henrich, S.~Stebner, S.~M{\"u}nstermann, A novel genetic algorithm-based calibration framework for crystal plasticity parameters in dp780 steels using multiscale mechanical testing, Computational Materials Science 258 (2025) 114088.
    \newblock \href {https://doi.org/10.1016/j.commatsci.2025.114088} {\path{doi:10.1016/j.commatsci.2025.114088}}.
    
    \bibitem{Lee.2024}
    D.~Lee, W.~W. Chen, L.~Wang, Y.-C. Chan, W.~Chen, Data-driven design for metamaterials and multiscale systems: A review, Advanced materials (Deerfield Beach, Fla.) 36~(8) (2024) e2305254.
    \newblock \href {https://doi.org/10.1002/adma.202305254} {\path{doi:10.1002/adma.202305254}}.
    
    \bibitem{Keijzer.2004}
    M.~Keijzer, Scaled symbolic regression, Genetic Programming and Evolvable Machines 5~(3) (2004) 259--269.
    \newblock \href {https://doi.org/10.1023/B:GENP.0000030195.77571.f9} {\path{doi:10.1023/B:GENP.0000030195.77571.f9}}.
    
    \bibitem{Zhang.2024b}
    Z.~Zhang, Z.~Zou, E.~Kuhl, G.~E. Karniadakis, Discovering a reaction--diffusion model for alzheimer's disease by combining pinns with symbolic regression, Computer Methods in Applied Mechanics and Engineering 419 (2024) 116647.
    \newblock \href {https://doi.org/10.1016/j.cma.2023.116647} {\path{doi:10.1016/j.cma.2023.116647}}.
    
    \bibitem{Xu.2023b}
    P.~Xu, X.~Ji, M.~Li, W.~Lu, Small data machine learning in materials science, npj Computational Materials 9~(1) (2023) 1--15.
    \newblock \href {https://doi.org/10.1038/s41524-023-01000-z} {\path{doi:10.1038/s41524-023-01000-z}}.
    
    \bibitem{Zhang.2024d}
    P.~Zhang, K.~Tang, A.~Wang, H.~Wu, Z.~Zhong, Neural network integrated with symbolic regression for multiaxial fatigue life prediction, International Journal of Fatigue 188 (2024) 108535.
    \newblock \href {https://doi.org/10.1016/j.ijfatigue.2024.108535} {\path{doi:10.1016/j.ijfatigue.2024.108535}}.
    
    \bibitem{Raissi.2019}
    M.~Raissi, P.~Perdikaris, G.~E. Karniadakis, Physics-informed neural networks: A deep learning framework for solving forward and inverse problems involving nonlinear partial differential equations, Journal of Computational Physics 378 (2019) 686--707.
    \newblock \href {https://doi.org/10.1016/j.jcp.2018.10.045} {\path{doi:10.1016/j.jcp.2018.10.045}}.
    
    \bibitem{Xiong.2025}
    W.~Xiong, X.~Long, S.~P. Bordas, C.~Jiang, The deep finite element method: A deep learning framework integrating the physics-informed neural networks with the finite element method, Computer Methods in Applied Mechanics and Engineering 436 (2025) 117681.
    \newblock \href {https://doi.org/10.1016/j.cma.2024.117681} {\path{doi:10.1016/j.cma.2024.117681}}.
    
    \bibitem{Luo.2024}
    C.~Luo, S.~Zhu, B.~Keshtegar, W.~Macek, R.~Branco, D.~Meng, Active kriging-based conjugate first-order reliability method for highly efficient structural reliability analysis using resample strategy, Computer Methods in Applied Mechanics and Engineering 423 (2024) 116863.
    \newblock \href {https://doi.org/10.1016/j.cma.2024.116863} {\path{doi:10.1016/j.cma.2024.116863}}.
    
    \bibitem{Weng.2025}
    H.~Weng, F.~Bamer, C.~Luo, B.~Markert, H.~Yuan, Physics-informed neural network for constitutive modeling of cyclic crystal plasticity considering deformation mechanism, International Journal of Mechanical Sciences 302 (2025) 110491.
    \newblock \href {https://doi.org/10.1016/j.ijmecsci.2025.110491} {\path{doi:10.1016/j.ijmecsci.2025.110491}}.
    
    \bibitem{Hu.2024c}
    H.~Hu, L.~Qi, X.~Chao, Physics-informed neural networks (pinn) for computational solid mechanics: Numerical frameworks and applications, Thin-Walled Structures 205 (2024) 112495.
    \newblock \href {https://doi.org/10.1016/j.tws.2024.112495} {\path{doi:10.1016/j.tws.2024.112495}}.
    
    \bibitem{Jiang.2024}
    L.~Jiang, Y.~Hu, Y.~Liu, X.~Zhang, G.~Kang, Q.~Kan, Physics-informed machine learning for low-cycle fatigue life prediction of 316 stainless steels, International Journal of Fatigue 182 (2024) 108187.
    \newblock \href {https://doi.org/10.1016/j.ijfatigue.2024.108187} {\path{doi:10.1016/j.ijfatigue.2024.108187}}.
    
    \bibitem{Feng.2024}
    F.~Feng, T.~Zhu, B.~Yang, S.~Zhou, S.~Xiao, A physics-informed neural network approach for predicting fatigue life of slm 316l stainless steel based on defect features, International Journal of Fatigue 188 (2024) 108486.
    \newblock \href {https://doi.org/10.1016/j.ijfatigue.2024.108486} {\path{doi:10.1016/j.ijfatigue.2024.108486}}.
    
    \bibitem{Feng.2025}
    F.~Feng, T.~Zhu, B.~Yang, Z.~Zhang, S.~Zhou, S.~Xiao, Probabilistic fatigue life prediction in additive manufacturing materials with a physics-informed neural network framework, Expert Systems with Applications 275 (2025) 127098.
    \newblock \href {https://doi.org/10.1016/j.eswa.2025.127098} {\path{doi:10.1016/j.eswa.2025.127098}}.
    
    \bibitem{Zhou.2025}
    S.~Zhou, M.~Henrich, Z.~Wei, F.~Feng, B.~Yang, S.~M{\"u}nstermann, A general physics-informed neural network framework for fatigue life prediction of metallic materials, Engineering Fracture Mechanics 322 (2025) 111136.
    \newblock \href {https://doi.org/10.1016/j.engfracmech.2025.111136} {\path{doi:10.1016/j.engfracmech.2025.111136}}.
    
    \bibitem{Goswami.2020}
    S.~Goswami, C.~Anitescu, S.~Chakraborty, T.~Rabczuk, Transfer learning enhanced physics informed neural network for phase-field modeling of fracture, Theoretical and Applied Fracture Mechanics 106 (2020) 102447.
    \newblock \href {https://doi.org/10.1016/j.tafmec.2019.102447} {\path{doi:10.1016/j.tafmec.2019.102447}}.
    
    \bibitem{Goswami.2022}
    S.~Goswami, M.~Yin, Y.~Yu, G.~E. Karniadakis, A physics-informed variational deeponet for predicting crack path in quasi-brittle materials, Computer Methods in Applied Mechanics and Engineering 391 (2022) 114587.
    \newblock \href {https://doi.org/10.1016/j.cma.2022.114587} {\path{doi:10.1016/j.cma.2022.114587}}.
    
    \bibitem{Zheng.2022}
    B.~Zheng, T.~Li, H.~Qi, L.~Gao, X.~Liu, L.~Yuan, Physics-informed machine learning model for computational fracture of quasi-brittle materials without labelled data, International Journal of Mechanical Sciences 223 (2022) 107282.
    \newblock \href {https://doi.org/10.1016/j.ijmecsci.2022.107282} {\path{doi:10.1016/j.ijmecsci.2022.107282}}.
    
    \bibitem{Zhao.2025}
    L.~Zhao, Q.~Shao, Denns: Discontinuity-embedded neural networks for fracture mechanics, Computer Methods in Applied Mechanics and Engineering 446 (2025) 118184.
    \newblock \href {https://doi.org/10.1016/j.cma.2025.118184} {\path{doi:10.1016/j.cma.2025.118184}}.
    
    \bibitem{Kolditz.2024}
    L.~M. Kolditz, S.~Dray, V.~Kosin, A.~Fau, F.~Hild, T.~Wick, Employing williams' series for the identification of fracture mechanics parameters from phase-field simulations, Engineering Fracture Mechanics 307 (2024) 110298.
    \newblock \href {https://doi.org/10.1016/j.engfracmech.2024.110298} {\path{doi:10.1016/j.engfracmech.2024.110298}}.
    
    \bibitem{Gu.2023}
    Y.~Gu, C.~Zhang, P.~Zhang, M.~V. Golub, B.~Yu, Enriched physics-informed neural networks for 2d in-plane crack analysis: Theory and matlab code, International Journal of Solids and Structures 276 (2023) 112321.
    \newblock \href {https://doi.org/10.1016/j.ijsolstr.2023.112321} {\path{doi:10.1016/j.ijsolstr.2023.112321}}.
    
    \bibitem{Chen.2024}
    Z.~Chen, Y.~Dai, Y.~Liu, Crack propagation simulation and overload fatigue life prediction via enhanced physics-informed neural networks, International Journal of Fatigue 186 (2024) 108382.
    \newblock \href {https://doi.org/10.1016/j.ijfatigue.2024.108382} {\path{doi:10.1016/j.ijfatigue.2024.108382}}.
    
    \bibitem{Gu.2024}
    Y.~Gu, L.~Xie, W.~Qu, S.~Zhao, Interface crack analysis in 2d bounded dissimilar materials using an enriched physics-informed neural networks, Engineering Analysis with Boundary Elements 163 (2024) 465--473.
    \newblock \href {https://doi.org/10.1016/j.enganabound.2024.03.030} {\path{doi:10.1016/j.enganabound.2024.03.030}}.
    
    \bibitem{Calafa.2024}
    M.~Calaf{\`a}, E.~Hovad, A.~P. Engsig-Karup, T.~Andriollo, Physics-informed holomorphic neural networks (pihnns): Solving 2d linear elasticity problems, Computer Methods in Applied Mechanics and Engineering 432 (2024) 117406.
    \newblock \href {https://doi.org/10.1016/j.cma.2024.117406} {\path{doi:10.1016/j.cma.2024.117406}}.
    
    \bibitem{Calafa.2025}
    M.~Calaf{\`a}, H.~M. Jensen, T.~Andriollo, Solving plane crack problems via enriched holomorphic neural networks, Engineering Fracture Mechanics 322 (2025) 111133.
    \newblock \href {https://doi.org/10.1016/j.engfracmech.2025.111133} {\path{doi:10.1016/j.engfracmech.2025.111133}}.
    
    \bibitem{He.2015}
    K.~He, X.~Zhang, S.~Ren, J.~Sun, Delving deep into rectifiers: Surpassing human-level performance on imagenet classification, in: 2015 IEEE International Conference on Computer Vision (ICCV), IEEE, Santiago, Chile, 2015, pp. 1026--1034.
    \newblock \href {https://doi.org/10.1109/ICCV.2015.123} {\path{doi:10.1109/ICCV.2015.123}}.
    
    \bibitem{Wang.2024m}
    L.-X. Wang, L.-F. Wen, R.~Tian, C.~Feng, Improved xfem (ixfem): Arbitrary multiple crack initiation, propagation and interaction analysis, Computer Methods in Applied Mechanics and Engineering 421 (2024) 116791.
    \newblock \href {https://doi.org/10.1016/j.cma.2024.116791} {\path{doi:10.1016/j.cma.2024.116791}}.
    
    \bibitem{Wang.2024h}
    Y.~Wang, C.-Y. Lai, Multi-stage neural networks: Function approximator of machine precision, Journal of Computational Physics 504 (2024) 112865.
    \newblock \href {https://doi.org/10.1016/j.jcp.2024.112865} {\path{doi:10.1016/j.jcp.2024.112865}}.
    
    \bibitem{Kingma.2015}
    D.~Kingma, J.~Ba, Adam: A method for stochastic optimization, in: Proceedings of the 3rd International Conference on Learning Representations (ICLR), San Diego, CA, USA, 2015.
    
    \bibitem{Zhang.2026}
    Z.~Zhang, G.~Yuan, Z.~Qin, Q.~Luo, An improvement by introducing lbfgs idea into the adam optimizer for machine learning, Expert Systems with Applications 296 (2026) 129002.
    \newblock \href {https://doi.org/10.1016/j.eswa.2025.129002} {\path{doi:10.1016/j.eswa.2025.129002}}.
    
    \bibitem{Abdolvand.2022}
    H.~Abdolvand, Development of microstructure-sensitive damage models for zirconium polycrystals, International Journal of Plasticity 149 (2022) 103156.
    \newblock \href {https://doi.org/10.1016/j.ijplas.2021.103156} {\path{doi:10.1016/j.ijplas.2021.103156}}.
    
    \bibitem{Rice.1968}
    J.~R. Rice, Mathematical analysis in the mechanics of fracture, Fracture: an advanced treatise 2 (1968) 191--311.
    
    \bibitem{Yau.1980}
    J.~F. Yau, S.~S. Wang, H.~T. Corten, A mixed-mode crack analysis of isotropic solids using conservation laws of elasticity, Journal of Applied Mechanics 47 (1980) 335--341.
    
    \bibitem{Shih.1986}
    C.~F. Shih, B.~Moran, T.~Nakamura, Energy release rate along a three-dimensional crack front in a thermally stressed body, International Journal of Fracture 30~(2) (1986) 79--102.
    \newblock \href {https://doi.org/10.1007/BF00034019} {\path{doi:10.1007/BF00034019}}.
    
    \bibitem{Paik.2021}
    S.~Paik, B.~K. Dutta, N.~N. Kumar, R.~Tewari, Fracture initiation in a single crystal copper edge-crack specimen for various crystallographic orientations, Theoretical and Applied Fracture Mechanics 114 (2021) 103019.
    \newblock \href {https://doi.org/10.1016/j.tafmec.2021.103019} {\path{doi:10.1016/j.tafmec.2021.103019}}.
    
    \bibitem{Erdogan.1963}
    F.~Erdogan, G.~Sih, On the crack extension in plates under plane loading and transverse shear, Journal of basic engineering 85~(4) (1963) 519--525.
    
    \bibitem{Zhao.2024c}
    Y.~Zhao, K.~Zheng, C.~Wang, Rock Fracture Mechanics and Fracture Criteria, 1st Edition, {Springer Nature Singapore} and {Imprint Springer}, Singapore, 2024.
    \newblock \href {https://doi.org/10.1007/978-981-97-5822-7} {\path{doi:10.1007/978-981-97-5822-7}}.
    
    \bibitem{Ortellado.2025}
    L.~Ortellado, A.~Abate, A.~Santarossa, L.~R. G{\'o}mez, T.~P{\"o}schel, Principle of local symmetry in mixed-mode fracture, Communications Physics 8~(1) (2025).
    \newblock \href {https://doi.org/10.1038/s42005-025-02151-9} {\path{doi:10.1038/s42005-025-02151-9}}.
    
    \bibitem{Paris.1963}
    P.~Paris, F.~Erdogan, A critical analysis of crack propagation laws, Journal of Basic Engineering 85~(4) (1963) 528--533.
    \newblock \href {https://doi.org/10.1115/1.3656900} {\path{doi:10.1115/1.3656900}}.
    
    \bibitem{Zhu.2025b}
    Y.~Zhu, L.~Zhang, J.~Gao, Y.~Pan, Z.~Barsoum, W.~Dou, A transfer learning-based adaptive neural network material modeling framework, International Journal of Mechanical Sciences 305 (2025) 110757.
    \newblock \href {https://doi.org/10.1016/j.ijmecsci.2025.110757} {\path{doi:10.1016/j.ijmecsci.2025.110757}}.
    
    \bibitem{Cotterell.1980}
    B.~Cotterell, J.~R. Rice, Slightly curved or kinked cracks, International Journal of Fracture 16~(2) (1980) 155--169.
    \newblock \href {https://doi.org/10.1007/BF00012619} {\path{doi:10.1007/BF00012619}}.
    
    \bibitem{Si.2024}
    Y.~Si, Y.~Wei, Semi-analytical solutions of kinked edge cracks, Engineering Fracture Mechanics 309 (2024) 110392.
    \newblock \href {https://doi.org/10.1016/j.engfracmech.2024.110392} {\path{doi:10.1016/j.engfracmech.2024.110392}}.
    
    \bibitem{Mitchell.2017}
    N.~P. Mitchell, V.~Koning, V.~Vitelli, W.~T.~M. Irvine, Fracture in sheets draped on curved surfaces, Nature Materials 16~(1) (2017) 89--93.
    \newblock \href {https://doi.org/10.1038/nmat4733} {\path{doi:10.1038/nmat4733}}.
    
    \bibitem{Tada.1973}
    H.~Tada, P.~C. Paris, G.~R. Irwin, The stress analysis of cracks, Handbook, Del Research Corporation 34~(1973) (1973).
    
    \bibitem{Wang.2024l}
    S.~Wang, B.~Yang, S.~Zhou, J.~Li, S.~Xiao, Closure effect of i + ii mixed-mode crack for ea4t axle steel, Chinese Journal of Mechanical Engineering 37~(1) (2024).
    \newblock \href {https://doi.org/10.1186/s10033-024-01061-1} {\path{doi:10.1186/s10033-024-01061-1}}.
    
    \bibitem{Sakha.2022}
    M.~Sakha, M.~Nejati, A.~Aminzadeh, S.~Ghouli, M.~O. Saar, T.~Driesner, On the validation of mixed-mode i/ii crack growth theories for anisotropic rocks, International Journal of Solids and Structures 241 (2022) 111484.
    \newblock \href {https://doi.org/10.1016/j.ijsolstr.2022.111484} {\path{doi:10.1016/j.ijsolstr.2022.111484}}.
    
    \bibitem{Aliha.2016}
    M.~Aliha, A.~Bahmani, S.~Akhondi, Mixed mode fracture toughness testing of pmma with different three-point bend type specimens, European Journal of Mechanics - A/Solids 58 (2016) 148--162.
    \newblock \href {https://doi.org/10.1016/j.euromechsol.2016.01.012} {\path{doi:10.1016/j.euromechsol.2016.01.012}}.
    
    \bibitem{Chen.2022b}
    H.~Chen, H.~Xing, H.~Imtiaz, B.~Liu, How to obtain a more accurate maximum energy release rate for mixed mode fracture, Forces in Mechanics 7 (2022) 100077.
    \newblock \href {https://doi.org/10.1016/j.finmec.2022.100077} {\path{doi:10.1016/j.finmec.2022.100077}}.
    
    \bibitem{Hua.2023}
    W.~Hua, J.~Li, Z.~Zhu, A.~Li, J.~Huang, Z.~Gan, S.~Dong, A review of mixed mode i-ii fracture criteria and their applications in brittle or quasi-brittle fracture analysis, Theoretical and Applied Fracture Mechanics 124 (2023) 103741.
    \newblock \href {https://doi.org/10.1016/j.tafmec.2022.103741} {\path{doi:10.1016/j.tafmec.2022.103741}}.
    
    \bibitem{Zong.2026}
    Y.~Zong, A.~M. Tartakovsky, Mathematics of digital twins and transfer learning for systems governed by pde models, Computer Methods in Applied Mechanics and Engineering 448 (2026) 118450.
    \newblock \href {https://doi.org/10.1016/j.cma.2025.118450} {\path{doi:10.1016/j.cma.2025.118450}}.
    
    \bibitem{Shen.2022}
    F.~Shen, S.~M{\"u}nstermann, J.~Lian, A unified fracture criterion considering stress state dependent transition of failure mechanisms in bcc steels at --196 °c, International Journal of Plasticity 156 (2022) 103365.
    \newblock \href {https://doi.org/10.1016/j.ijplas.2022.103365} {\path{doi:10.1016/j.ijplas.2022.103365}}.
    
    \bibitem{James.2013}
    M.~N. James, C.~J. Christopher, Y.~Lu, E.~A. Patterson, Local crack plasticity and its influences on the global elastic stress field, International Journal of Fatigue 46 (2013) 4--15.
    \newblock \href {https://doi.org/10.1016/j.ijfatigue.2012.04.015} {\path{doi:10.1016/j.ijfatigue.2012.04.015}}.
    
\end{thebibliography}

\end{document}